\documentclass{emulateapj}
\begin{document}
\slugcomment{{\it The Astronomical Journal, accepted}}

\title{Lenticular galaxies at the outskirts of the Leo II group: NGC 3599 and NGC 3626
\footnote{Partly based on observations collected with the 6m
telescope at the Special Astrophysical Observatory (SAO) of the
Russian Academy of Sciences (RAS).}}

\author{O. K. Sil'chenko\altaffilmark{1}}
\affil{Sternberg Astronomical Institute, Moscow, 119992 Russia}
       \affil{Isaac Newton Institute, Chile, Moscow Branch}
     \affil{Electronic mail: olga@sai.msu.su}
\altaffiltext{1}{Guest Investigator of the UK Astronomy Data Centre}

\author{A. V. Moiseev}

\affil{Special Astrophysical Observatory, Nizhnij Arkhyz,
    369167 Russia}
\affil{Electronic mail: moisav@gmail.com}
 
\author{A. P. Shulga}

\affil{Sternberg Astronomical Institute, Moscow, 119992 Russia\\
   Electronic mail: alina.shulga@gmail.com}

\begin{abstract}

We have studied unbarred S0 galaxies, NGC~3599 and NGC~3626, the members
of the X-ray bright group Leo II, by means of 3D spectroscopy, long-slit
spectroscopy, and imaging, with the aim to identify epoch and mechanisms
of their transformation from spirals. Both galaxies have appeared to bear
a complex of features resulting obviously from minor merging:
decoupled gas kinematics, nuclear starforming rings, and multi-tiered
oval large-scale stellar disks. The weak-emission line nucleus of NGC 3599
bears all signs of the Seyfert activity, according to the line-ratio
diagnostics of the gas excitation mechanism. After all,
we conclude that the transformation of these lenticular galaxies has
had place about 1--2 Gyr ago, through the gravitational mechanisms
not related to hot intragroup medium of Leo II.

\end{abstract}

\keywords{galaxies: individual (NGC3599) ---
galaxies: individual (NGC3626) --- galaxies: evolution ---
galaxies: structure}

\section{Introduction}

Lenticular galaxies as a class are perhaps shaped rather recently. 
Observations of clusters at intermediate redshifts, at $z=0.2-0.7$,
catch the moment when the dominance of spiral galaxies in clusters is
replaced by the dominance of S0s \citep{dressler,fasano};
so it allows to state that lenticular galaxies as a type have been
formed near the redshift of $z\approx 0.3-0.4$, or 3-5 Gyr ago. The mechanism
of the transformation of spirals into S0s is, however, still uncertain.
The common point of view is that environment is a key point, but what
properties of environment -- hot medium, dense settlement, high pairwise
velocities, or something else -- play a role, is a topic of discussions.
Last years an idea has arisen that environment where S0s form is indeed
groups of galaxies; and later the groups bring S0s into the clusters when
accreting onto them \citep{wilman,just2010}. If so, just 
lenticulars in groups represent a particular interest for identifying the 
mechanisms of their (trans-)formation.

Nearby groups are very inhomogeneous as concerning their properties.
They may be dominated by late-type or by early-type galaxies, half of them
contain hot X-ray gas and half of them lack it \citep{xraygroup}; and the observed
ranges of group masses and of galaxy velocity dispersions are also rather
large -- from $10^{12}$ to $10^{15} \, M_{\odot}$ and from 50 to 800 km/s,
respectively \citep{catgroup}.
So study of lenticular galaxies in nearby groups with a spread of properties
may allow to identify the most important drivers of their evolution -- or to
confirm a feeling that there exist different paths to transform a spiral galaxy
into lenticular one.
We have already got some interesting results on this way. Lenticular galaxies
in groups reveal serious inhomogeneity of their properties. In the groups
NGC~5576 \citep{sil97,s0_3} and Leo I \citep{we2003}
where X-ray intragroup gas is absent the early-type galaxies demonstrate
synchronous nuclear evolution and have rather young, of 2-3 Gyrs old, nuclear
stellar populations; perhaps, this synchronity is a manifestation of the major
role of recent gravitational interactions. In the group NGC~80 
\citep{n80mpfs,n80phot} which looks quite relaxed and has
a prominent X-ray gaseous halo, the centers of some lenticular galaxies
are old (NGC~80, IC~1541, NGC~86) while the others (IC~1548, NGC~85) are
young and seem to (trans-)form quite recently due to minor mergers. Though
for several elder lenticulars we cannot exclude the major role of
hydrodynamical mechanisms, like ram pressure, the presence of `old'
lenticulars at the outskirts of the group and of the `younger' ones in the
center inverse the whole picture of the galaxy assembly.

The group Leo II is in some sense a middle case: it possesses X-ray gas, but the
whole structure of its X-ray halo is not very regular, with two prominent peaks
at two central galaxies, the S0 NGC~3607, the brightest galaxy in the group, and
at its elliptical neighbor, NGC~3608 \citep{xrayleo2}. We studied the 
central galaxies of the group \citep{leo2cen} and found the
different central ages of stellar populations, very old in NGC~3607 and 
intermediate one in NGC~3608, so in this group we saw no synchronity of the 
evolution. In this paper we continue to analyze structure and properties of 
the nuclear stellar populations of the group members, this time -- in two 
off-center lenticular galaxies of the group, NGC~3599 and NGC~3626. The global 
characteristics of the galaxies under consideration are listed in Table 1. 
NGC~3626 is famous by its counterrrotating large-scale gaseous disk
\citep{n3626ciri,n3626co,n3626hi}. 
The dwarf S0 NGC~3599 is not so famous; however recently a mysterious 
X-ray flash in its center has provoked a statement about a supermassive 
black hole having disrupted tidally a star \citep{n3599xray}, though 
the galaxy has not been reported to have a noticeable active nucleus.

\begin{table}
\caption[ ] {Global parameters of the galaxies}
% %\begin{center}
\begin{flushleft}
\begin{tabular}{lcc}
\hline\noalign{\smallskip}
% % & & & \\
NGC & 3599 & 3626  \\
\hline
Type (NED$^1$) & SA0: & (R)SA(rs)0$+$   \\
$R_{25}$, kpc (NED) & 7.1 & 7.8 \\
$B_T^0$ (RC3$^2$) & 12.70 & 11.62  \\
$M_B$(RC3$+$NED)  & --18.8 & --19.9  \\
$(B-V)_T^0$ (RC3) & 0.86 & 0.81  \\
$(U-B)_T^0$ (RC3) & 0.32 & 0.30 \\
$V_r $ (NED), $\mbox{km} \cdot \mbox{s}^{-1}$ &   832 &
1493   \\
Distance, Mpc  & $20.4^4$ & $20.0^4$ \\
Scale, pc/arcsec&  99 & 97 \\
Inclination (LEDA$^3$) & $28^{\circ}$ & $56^{\circ}$  \\
{\it PA}$_{phot}$ (LEDA) & -- & $156^{\circ}$  \\
$\sigma _*$, $\mbox{km} \cdot \mbox{s}^{-1}$(LEDA) & 67 & 142 \\
\hline
\multicolumn{3}{l}{$^1$\rule{0pt}{11pt}\footnotesize
NASA/IPAC Extragalactic Database}\\
\multicolumn{3}{l}{$^2$\rule{0pt}{11pt}\footnotesize
Third Reference Catalogue of Bright Galaxies}\\
\multicolumn{3}{l}{$^3$\rule{0pt}{11pt}\footnotesize
Lyon-Meudon Extragalactic Database}\\
\multicolumn{3}{l}{$^4$\rule{0pt}{11pt}\footnotesize
\citet{sbfdist}}
\end{tabular}
\end{flushleft}
\end{table}

\section{The observations and data used}

Our study of the lenticular galaxies NGC~3599 and NGC~3626 presented here
includes analysis of the structure of the galaxies and determination
of the characteristics of the central stellar populations. The former
aim is achieved by photometric decomposition of the SDSS images and by
comparing the photometric and kinematical orientation angles. The latter
piece of work is made by means of panoramic spectroscopy and by analysis
of the color maps. The list of the raw observational data used is given in Table 2.

Photometric data which we analyse include also $gri$-images from the DR6
of the Sloan Digital Sky Survey (SDSS) \citep{sdssdr6} 
for both galaxies, the HST two-color data for the central region of NGC~3599
(HST/WFPC2, the bands of F814W and F555W, Prop. no. 5999, PI. A. Phillips), 
together with our own $BV$-photometry for NGC~3626 undertaken with the focal reductor 
SCORPIO \citep{scorpref} of the Russian 6m telescope Big Telescope 
Alt-azimuthal (BTA) in the Special Astrophysical Observatory (SAO RAS).

Also, the focal reducer SCORPIO in the long-slit mode has been used to obtain
a red-range spectrum of NGC 3599. We hoped to probe the radial extension
of the ionized-gas emission in this galaxy, but the emission was found 
to be confined only to the very central part of the galaxy, $R<10\arcsec $.

The panoramic spectroscopy was made with the integral-field spectrograph of the
6m telescope MPFS (`Multi-Pupil Fiber Spectrograph') \citep{mpfsref};
also we have involved the data from the William Herschel Telescope integral-field
spectrograph SAURON \citep{betal01} retrieved from the open Isaac Newton
Group Archive which is maintained as part of the CASU Astronomical Data
Centre at the Institute of Astronomy, Cambridge.

Integral-field spectroscopy provides simultaneous exposing wide-range spectra
for  every spatial element (spaxel) from a two-dimensional area on the sky.
So, by observing galaxies with the integral-field spectrograph, we get a
possibility of mapping various spectral characteristics of the light integrated
on the line of sight: surface brightness, both for the stellar continuum and
gas emission lines, line-of-sight velocities and velocity dispersions (or,
in more general sense, line-of-sight velocity distribution, LOSVD, shape
parameters), and equivalent widths of absorption lines which we express
in the well-formulated Lick index system \citep{woretal}. The MPFS is
operating on the fiber-lens principle, so it provides a large free spectral
range, about 1500~\AA\ with the spectral resolution of 3~\AA\
by using the CCD EEV 42--40 of $2048 \times 2048$ format as a detector;
the price is a small field of view, $16\times 16$ spaxels, with the sampling
of $1\arcsec$ per spaxel. The SAURON operates in TIGER mode \citep{betal95}
so it exposes a fixed narrow spectral range, 4800--5350~\AA, but packs densely
more than 1500 spectra covering $44\times 38$ spaxels. The `blank sky' spectra
are exposed simultaneously with the galaxy spectra, on the same detector; they are 
taken at $4\arcmin$ from the field center by the MPFS and at $1.7\arcmin$ from the
field center by the SAURON. After the necessary primary reduction steps -- bias
subtraction, flatfielding, individual spectra extraction, their calibration
onto the wavelength scale, sky subtraction -- we analyse every spectrum
individually, and after deriving interesting spectral characteristics,
we combine them into the maps of the galaxy regions studied. The star velocities
and stellar velocity dispersions are calculated by cross-correlating galaxy
spectra with the spectra of G8--K2-giant stars observed the same nights, and
with the spectra of twilight sky which is of G2-type. Several templates are
taken to estimate sensitivity of the velocities derived to the possible
template mismatch; it has appeared to be negligible under our approach. The gas 
velocities are measured by calculating the baricenter positions of emission lines. 
The Lick indices H$\beta$, Mgb, Fe5270, and Fe5335 are calculated from the
individual galaxy spectra, corrected for the stellar velocity dispersion,
and calibrated into the standard Lick system by using the observations
of a dozen Lick standard stars \citep{woretal} observed the same
observational run as the galaxies -- for the procedure in detail see e.g. 
\citet{lenssum} or \citet{leo2cen}. The typical errors of the individual velocity
measurements are of 10 km/s, the errors of the Lick indices vary between
0.1~\AA\ in the centers of galaxies to 0.5~\AA\ at the edges of the field
of view.

\begin{table*}
\scriptsize
\caption[ ] {Spectroscopy and photometry of the galaxies studied}
% \begin{center}
\begin{flushleft}
\begin{tabular}{lllllccc}
\hline\noalign{\smallskip}
Date & Galaxy & Exposure & Spectrograph & Field & PA(top) &
Spectral range & Seeing \\
\hline\noalign{\smallskip}
13 Apr 05 & NGC~3599 & 100 min & BTA/MPFS &
$16\arcsec \times 16\arcsec $ & $154.5^{\circ}$ & 4200-5600~\AA\ & $1\farcs 5$ \\
05 Apr 06 & NGC~3599 & 60 min & BTA/MPFS &
$16\arcsec \times 16\arcsec $ & $276^{\circ}$ & 5800-7200~\AA\ & $2\farcs 0$ \\
14 Apr 05 & NGC~3626 &  60 min & BTA/MPFS &
$16\arcsec \times 16\arcsec $ & $260.5^{\circ}$ & 4200-5600~\AA\ & $1\farcs 5$ \\
05 Apr 06 & NGC~3626 &  30 min & BTA/MPFS &
$16\arcsec \times 16\arcsec $ & $261^{\circ}$ & 5800-7200~\AA\ & $2\farcs 0$ \\
09 Jan 08 & NGC~3626 & 60 min & WHT/SAURON  &
$36\arcsec\times 41\arcsec$ & $155^{\circ}$ & 4800-5350~\AA\ & $1\farcs 7$ \\
29 Feb 08 & NGC~3599 & 60 min & WHT/SAURON  &
$36\arcsec\times 41\arcsec$ & $215^{\circ}$ & 4800-5350~\AA\ & $1\farcs 1$ \\
05 Apr 09 & NGC~3599 & 30 min & BTA/SCORPIO/LS &
$1\arcsec \times 360\arcsec $ & $240^{\circ}$ & 6100-7100~\AA\ & $1\farcs 3$ \\
25 Dec 05 & NGC~3626 & 4 min & BTA/SCORPIO/IM &
$360\arcsec \times 360\arcsec $ & $0^{\circ}$ & $B$-band & $1\farcs 8$ \\
25 Dec 05 & NGC~3626 & 6 min & BTA/SCORPIO/IM &
$360\arcsec \times 360\arcsec $ & $0^{\circ}$ & $V$-band & $1\farcs 7$ \\
\hline
\end{tabular}
% \end{center}
\end{flushleft}
\end{table*}

\section{The large-scale structure of NGC 3599 and NGC 3626}

We have analysed the large-scale structure of the galaxies by applying
the 2D decomposition software BUDDA \citep{budda} to the images taken
from the SDSS DR6 and then sky-subtracted. The results of decomposition
in three bands, $g$, $r$, and $i$, are given in the Table 3. The present version
of the BUDDA allows to decompose a galaxy into exponential disk and a bulge
describing by the Sersic profile with an arbitrary power parameter.
However we know that an exponential
scalelength may vary along the radius -- disks may be two-tiered, truncated
or antitruncated \citep{sdsslate,erwinbars}. To catch this
peculiarity of the structures, we apply a more complex approach to the galaxy
decompositions. Basing on isophotal analysis, we define an outer zone where
the outer disk dominates -- it is the zones where the isophote PA and ellipticity
are constant along the radius, -- mask the inner part of the galaxy, and for the first
time apply the BUDDA procedure only to the outer disk-dominated zone. After
obtaining the parameters of the outer disk, we subtract the full-extension model 
disk image from the original image of the galaxy and then apply the BUDDA to the 
residual images to derive the parameters of the inner disk and of the bulge. 
For NGC~3599 we have been forced to repeat this procedure: this galaxy has 
appeared to possess a three-tiered disk, with the middle scalelength being 
the longest. The isophote analysis has been made in the framework of IRAF.

Figure~1 and Figure~2 present the results of the isophote analysis in one of the filters
for NGC~3599 and NGC~3626, correspondingly (the isophote behaviors in all filters
are similar): the left plots refer to the galaxies as the wholes and the right plots 
-- to the galaxies with the outer disks subtracted. Figures 1-left and 2-left demonstrate
a common feature: there exists a transition radius where the major axis position
angles change abruptly and the ellipticities drop. At the same radii, the 
disk profiles have breaks of the slopes. The isophote-analysis 
results for the surface brightness distribution in NGC~3599 consistent with those
presented in Fig.~1 have been obtained by \citet{magrelli} in the $V$-band.  
Here we place the isophote-analysis results for NGC~3599 
and NGC~3626 together to point out the similarity of their structures.

We suggest that in Figs.~1 and 2 we see transitions between the inner and 
outer disks (as for the third disk in NGC~3599 we can treat it as a truncation 
portion of the outer disk because it has the same PA and ellipticity and 
differs only by the scalelength). We propose the following arguments for 
treating the inner exponential structures with the high isophote ellipticities 
as quite separate inner disks and not as bars within the single-exponential
(outer) disks. Firstly, within the inner-disk zones the isophotes demonstrate
nearly constant high ellipticities, with low-contrast maxima at the inner edges
corresponding to the possible stellar rings (Figs. 1-left and 2-left), while bars,
having flat brightness profiles in early-type galaxies \citep{eebars85}, 
being superposed onto the exponential-profile disks would force the ellipticities 
to rise constantly along the radius towards the bars' ends \citep{3bphot}. Secondly, 
over the extension of the inner disks the surface brightness profiles demonstrate 
regular exponential shape with the scalelengths significantly smaller than those of the 
outer disks (the steeper slopes). In the barred single-exponential disks, the azimuthally 
averaged surface brightness profiles are very smooth, with a little, if any, influence 
of the ends of bars on the profile slopes \citep{eebars85,eebars96}. So we conclude 
that the photometric analysis implies a presence of the inner, separate 
stellar disks in both NGC~3599 and NGC~3626; in NGC~3599 it is a pure exponential-profile
structure (Fig.~1-right), while in NGC~3626 a sort of a flat-profile lens may also be
present (Fig.~2-right) -- see also the decomposition of NGC~3626 by \citet{lauri05}. 
If we assume that the outer isophote PAs and ellipticities relate to the orientation 
of the line of nodes and to the  cosine of inclination of the galaxy symmetry planes, 
respectively, -- and it is so, if the outer disks are round, -- then the inner disks 
must be oval, or round but inclined to the planes of the outer disks. To choose between 
these alternatives, we need kinematical data: the photometry alone cannot discriminate 
between these two configurations.

\begin{table*}

\scriptsize%{tiny}

\caption[ ] {Sersic parameters of the brightness profiles fitting}

% \begin{center}

\begin{flushleft}

\begin{tabular}{lccccccc}

\hline

Component & Radius range of fitting, arcsec & n

 & $PA_0$  & $1-b/a$ & $\mu_0$, mag/sq. arcsec &

$r_0$, arcsec & $r_e$, arcsec \\

\hline

\multicolumn{8}{l}{NGC~3599, i-band}\\

Outermost disk & $>60$ & 1 & $52\degr$ & $0.11\pm 0.05$ &

$20.6 \pm 1.0$ & $25 \pm 4$  & -- \\

Outer disk & 30--60 & 1 & $53\degr$ & $0.11\pm 0.05$ &

$20.8 \pm 0.4$ &  $28 \pm 4$ & -- \\

Inner disk & 10--35 & 1 & $101\degr$ & $0.20 \pm 0.01$ &

$19.5 \pm 0.1 $ & $14.0 \pm 0.6$ & -- \\

Central bulge & $<10$ & $4.28\pm 0.45$ &  $94\degr$ & $0.12 \pm 0.01$ & --

& -- & $3.18 \pm 0.08$ \\

\multicolumn{8}{l}{NGC~3599, r-band}\\

Outermost disk & $>60$ & 1 & $52\degr$ & $0.07\pm 0.04$ &

$20.6 \pm 1.0$ & $22 \pm 3$  & -- \\

Outer disk & 30--60 & 1 & $55\degr$ & $0.10\pm 0.05$ &

$21.4 \pm 0.5$ &  $30 \pm 4$ & -- \\

Inner disk & 10--35 & 1 & $101\degr$ & $0.23 \pm 0.03$ &

$19.9 \pm 0.1 $ & $13.2 \pm 0.7$ & -- \\

Central bulge & $<10$ & $4.24\pm 0.50$ &  $95\degr$ & $0.10 \pm 0.01$ & --

& -- & $3.14 \pm 0.11$ \\

\multicolumn{8}{l}{NGC~3599, g-band}\\

Outermost disk & $>60$ & 1 & $52\degr$ & $0.06\pm 0.03$ &

$21 \pm 1.0$ & $20 \pm 4$  & -- \\

Outer disk & 30--60 & 1 & $57\degr$ & $0.12\pm 0.00$ &

$22.0 \pm 0.6$ &  $30 \pm 5$ & -- \\

Inner disk & 10--35 & 1 & $102\degr$ & $0.22 \pm 0.03$ &

$20.4 \pm 0.2 $ & $12.6 \pm 0.7$ & -- \\

Central bulge & $<10$ & $3.75\pm 0.53$ &  $81\degr$ & $0.09 \pm 0.01$ & --

& -- & $2.77 \pm 0.11$ \\

\hline

\multicolumn{8}{l}{NGC~3626, i-band}\\

Outer disk & $>45$ & 1 & $159\degr$ & $0.31\pm 0.05$ &

$19.1 \pm 0.5$ &  $21 \pm 2$ & -- \\

Inner disk & 10--45 & 1 & $168\degr$ & $0.48 \pm 0.02$ &

$18.6 \pm 0.1 $ & $19.8 \pm 0.7$ & -- \\

Central bulge & $<10$ & $2.1\pm 0.1$ &  $164\degr$ & $0.32 \pm 0.02$ & --

& -- & $2.52 \pm 0.05$ \\

\multicolumn{8}{l}{NGC~3626, r-band}\\

Outer disk & $>45$ & 1 & $160\degr$ & $0.31\pm 0.05$ &

$19.5 \pm 0.5$ &  $21 \pm 2$ & -- \\

Inner disk & 10--45 & 1 & $168\degr$ & $0.49 \pm 0.02$ &

$19.0 \pm 0.1 $ & $19.8 \pm 0.6$ & -- \\

Central bulge & $<10$ & $2.07\pm 0.10$ &  $164\degr$ & $0.33 \pm 0.02$ & --

& -- & $2.47 \pm 0.06$ \\

\multicolumn{8}{l}{NGC~3626, g-band}\\

Outer disk & $>45$ & 1 & $164\degr$ & $0.32\pm 0.06$ &

$20.4 \pm 0.6$ &  $24 \pm 2$ & -- \\

Inner disk & 10--45 & 1 & $169\degr$ & $0.50 \pm 0.02$ &

$19.8 \pm 0.1 $ & $21.6 \pm 0.7$ & -- \\

Central bulge & $<10$ & $1.73\pm 0.21$ &  $163\degr$ & $0.35 \pm 0.02$ & --

& -- & $2.73 \pm 0.06$ \\

\hline

\end{tabular}

% \end{center}

\end{flushleft}

\end{table*}

\section{The kinematics of the central parts of NGC 3599 and NGC 3626.}

To compare the kinematics of the stars and of the ionized gas, we involve
both the data from the MPFS and from the SAURON. The SAURON field of view
is larger than that of the MPFS, but the SAURON observes only the green
spectral range where the strongest emission line, [\ion{O}{3}]$\lambda$5007
is still hardly measureable while sinking to the bottom of the prominent \ion{Ti}{1}
absorption line. The MPFS provides us with the measurements of the emission
line [\ion{N}{2}]$\lambda$6584 which is strong and free of underlying absorptions
and so gives us the most reliable velocity fields for the warm gas component.
The SAURON {\it stellar} velocity fields are preferred due to the larger field
of view. We apply a tilted-ring analysis to both gaseous and stellar LOS
velocity fields by using the software DETKA \citep{moiseevdb}, which estimates 
both $PA$ of the kinematical major axis and the inclination of the rotation plane.

\begin{figure*}
\plottwo{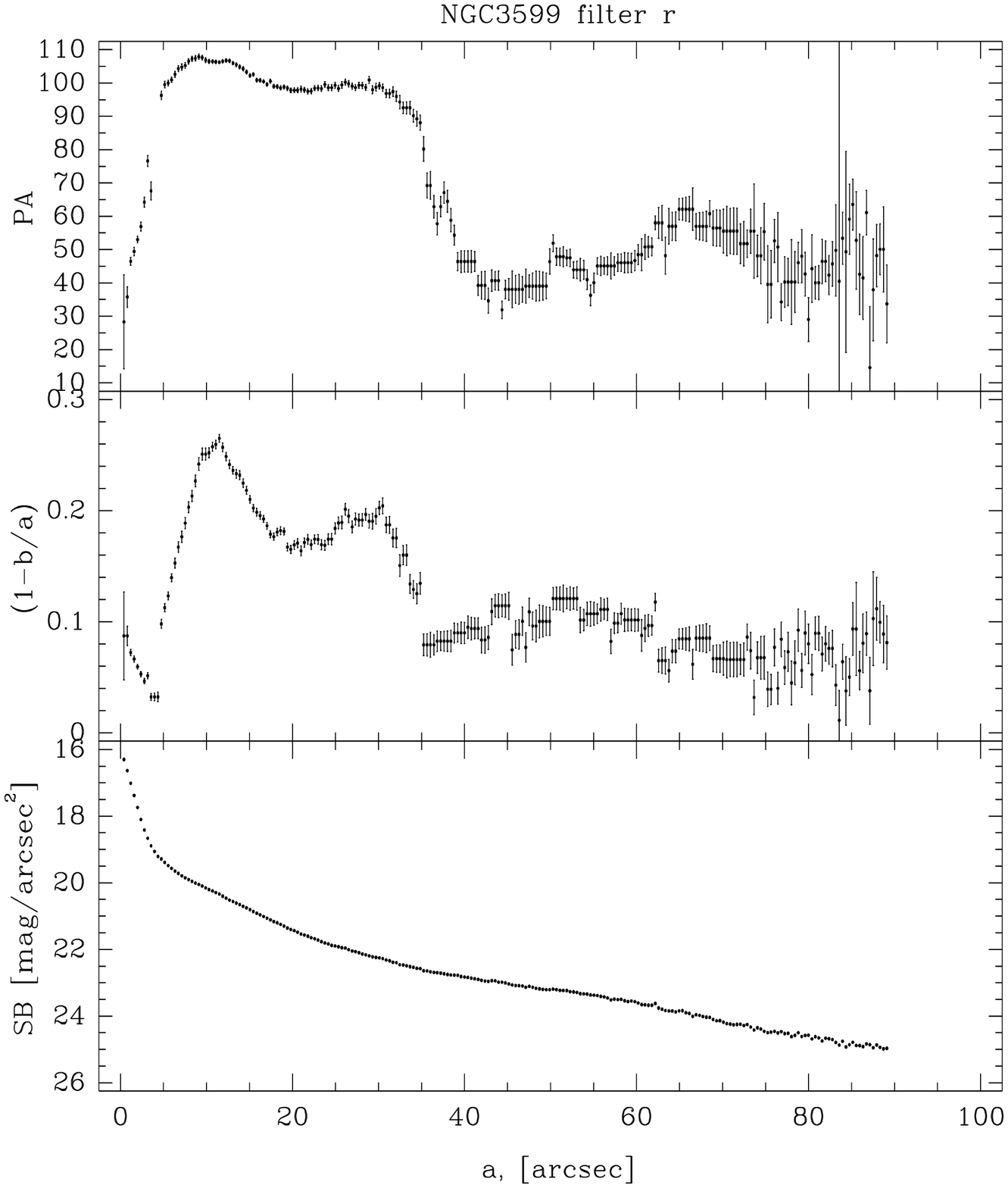}{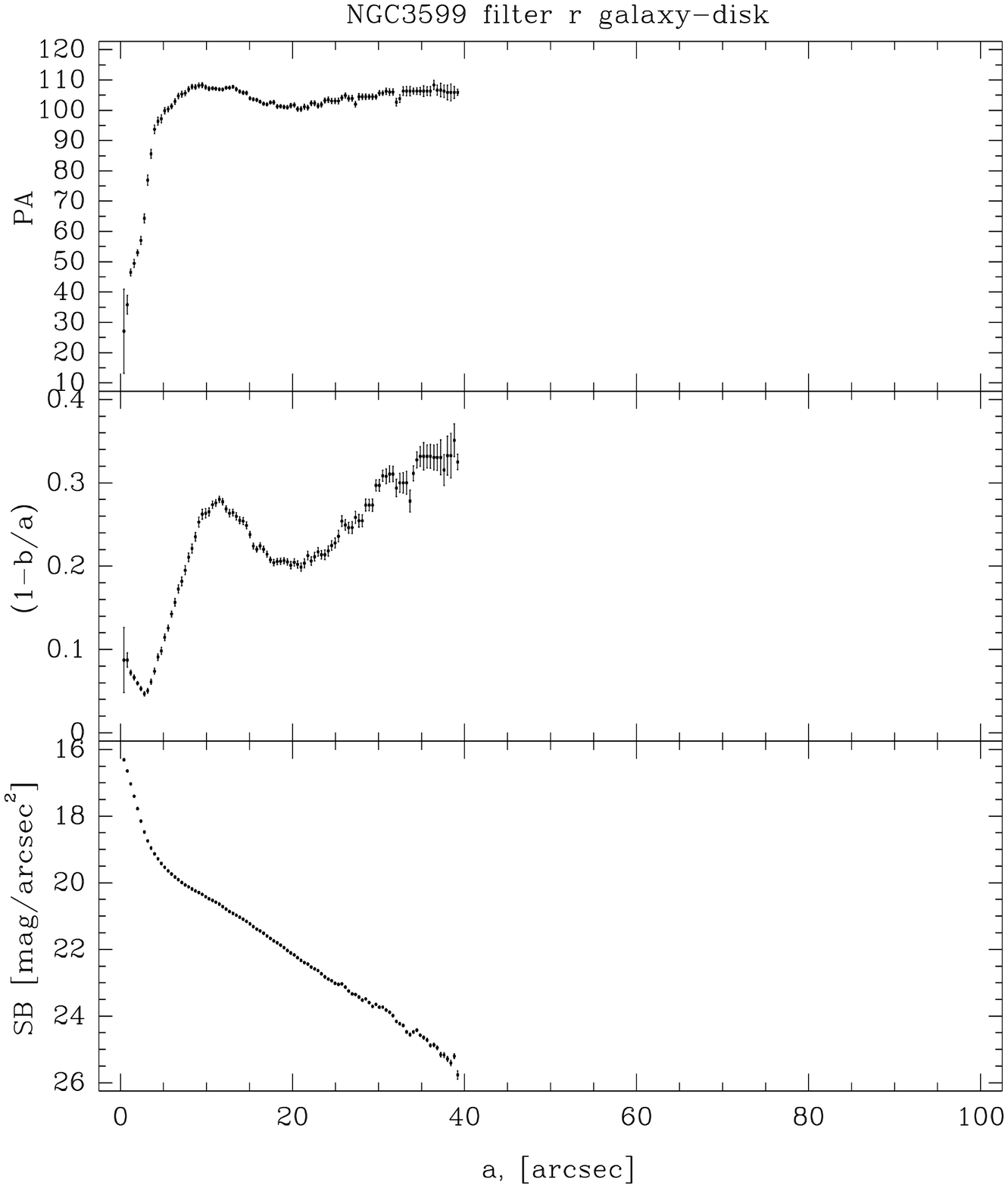}
\caption{The results of the isophotal analysis
for the $r$-band image of NGC~3599: {\bf left} -- for the full image,
{\bf right} -- for the residual image after subtracting the model outer disk.
In both plots
{\it top} -- the position angle of the isophote major axis, {\it middle} --
the ellipticity, {\it bottom} -- the azimuthally averaged surface brightness
profiles.}
\end{figure*}

\begin{figure*}
\plottwo{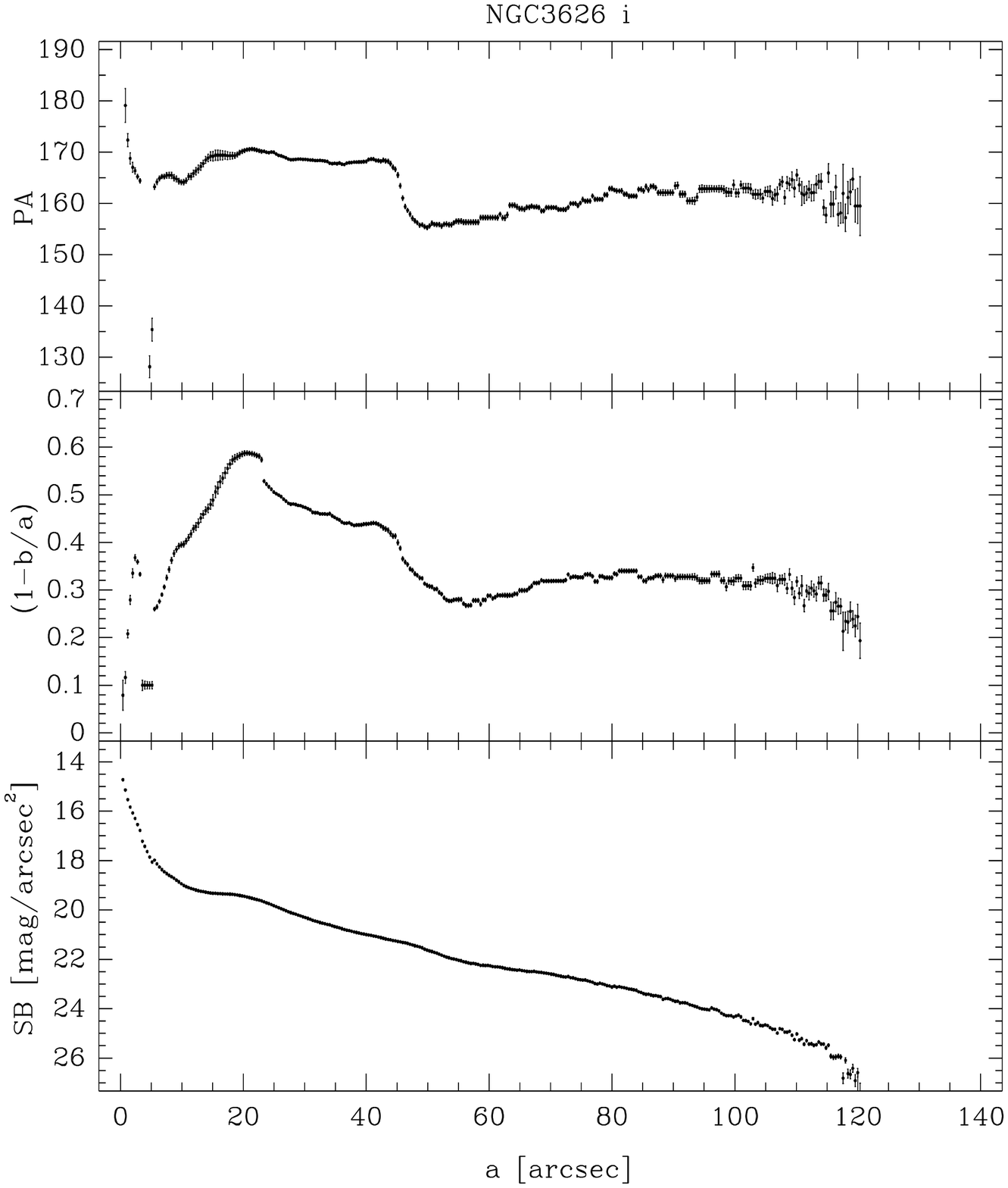}{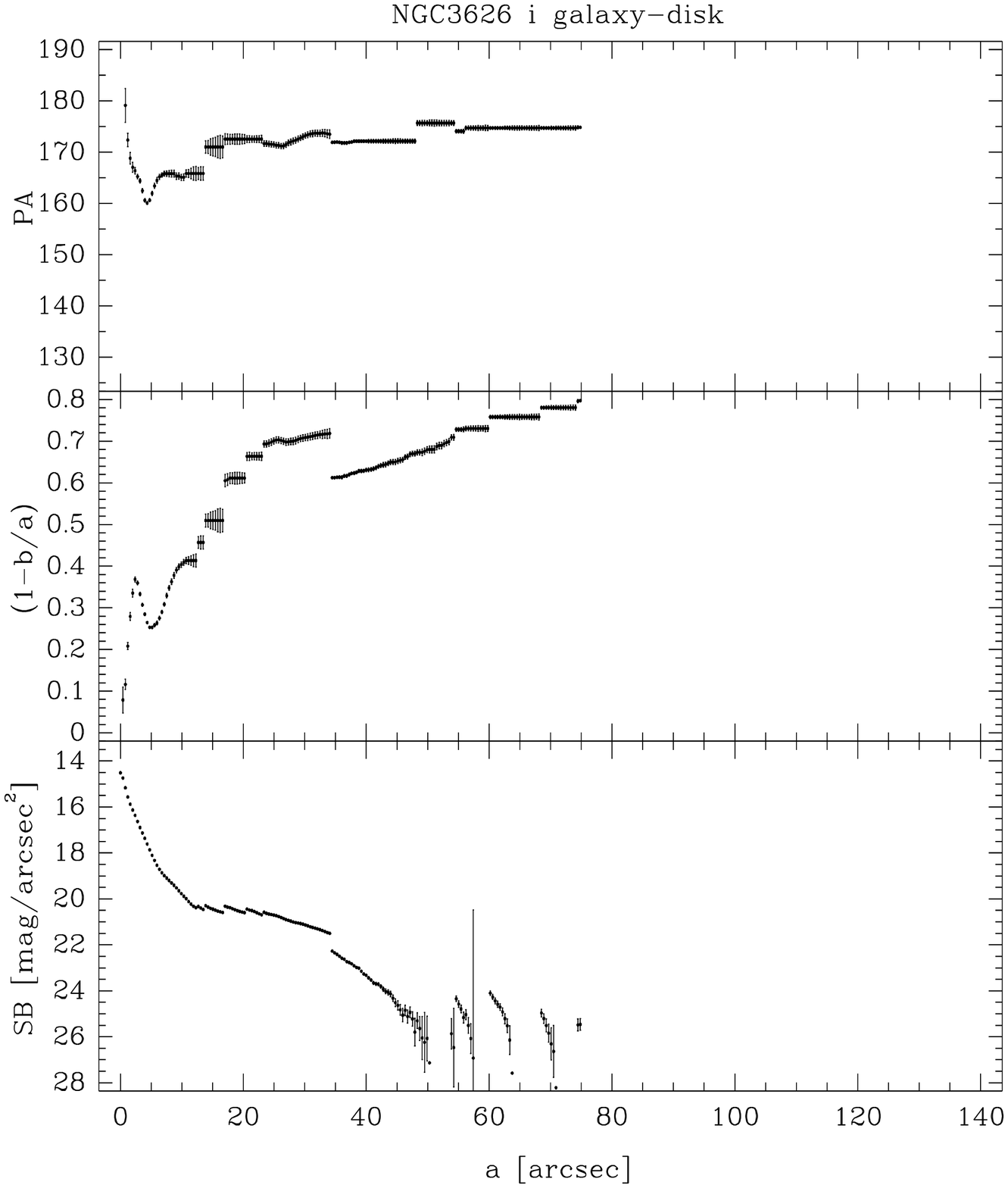}
\caption{The results of the isophotal analysis
for the $i$-band
image of NGC~3626: {\bf left} -- for the full image, {\bf right} -- for the 
residual image after subtracting the model outer disk. In both plots
{\it top} -- the position angle of the isophote major axis, {\it middle} --
the ellipticity, {\it bottom} -- the azimuthally averaged surface brightness
profiles.}
\end{figure*}

Figure~3 shows the line-of-sight velocity fields for the stars and for the
ionized gas in NGC~3599. The first visual inspection
gives evidence for the orthogonality of the stellar and gaseous rotation.
Indeed, while the kinematical major axis of the stellar component in the
center of the galaxy is aligned at $PA_0 (\mbox{stars})\approx 50\degr$,
the emission-line measurements both from the [\ion{N}{2}] and the [\ion{O}{3}]
data indicate $PA_0 (\mbox{gas}) \approx -30\degr - -50\degr$ at $R=2\arcsec-5\arcsec$.
By applying the DETKA to the SAURON stellar velocity field, we have obtained
radial  variations of the orientation angles up to $R=17\arcsec$ (Fig.~4).
There are no sure signs of the tilt of the stellar rotation plane. It is seen close
to face-on, with the inclination estimates being $28\pm 6\degr$ 
and the $PA_0 (kin)$ estimates -- between $45\degr$ and $65\degr$
(rather poorly restricted because of the close to face-on
rotation plane orientation). For the gas, we see a certain rotation-plane
tilt: it is close to the stellar rotation plane near the center and is
inclined by $\Delta i=50\degr \pm 10\degr$ to it at $R>6\arcsec$. The latter
configuration may be a nearly polar gas ring which warps to the galactic
plane near the very center.
In Fig.~4 we compare the orientations of the photometric and kinematical
major axes: it is the way to check an axisymmetry of the galaxy potential,
since within the axisymmetric potential the rotation must be circular, and
the kinematical and photometric major axes of round disks must coincide.
For NGC~3599, we see that the central part, $R<3\arcsec$, demonstrates
axisymmetric stellar rotation in the plane which line of nodes coincides with
the line of nodes of the outer disk (Table~3). In the inner disk area,
within $R=5\arcsec - 15\arcsec$, the stellar kinematical major axis oscillates
weakly near the line-of-nodes orientation while the isophotes turn by
some $50\degr$. With such a behavior of photometric and kinematical major axes
we can state that the inner disk in NGC~3599 is certainly oval.

Figure~5 presents the kinematical MPFS and SAURON maps for NGC~3626. The
large-scale gaseous disk in this galaxy is known to counterrotate the stars
\citep{n3626ciri,n3626hi}. We analyse here rotation, and other motions,
of the central gas within $R<17\arcsec$. In Fig.~5, right-bottom, we have masked
the regions of the ionized-gas velocity field where the intensity of the emission line 
[\ion{O}{3}]$\lambda$5007 is low, and so velocity measurements are uncertain. After 
that masking, the whole rotation of the highly excited warm gas can be 
separated into two distinct components: a counterrotating subsystem at 
$R>8\arcsec$ which is already known and an orthogonally projected gaseous ring 
at $R < 7\arcsec$ which has not been reported yet. We note an interesting
detail: the approaching `spot' of the ring is accompanied by the rise of
the stellar velocity dispersion (Fig.~5, right-top). The most natural explanation of
this local effect is a superposition of two stellar subsystems with slightly
different line-of-sight velocities. In other words, the inner gaseous ring
has the stellar counterpart. The previous studies of the gas kinematics in
the center of NGC~3626 gave already some hints on the multiple gas subsystems
within $R\approx 10\arcsec$. \citet{n3626hi} made two long-slit cross-sections --
along major ($PA=157\degr$) and minor ($PA=67\degr$) isophote axes of NGC~3626. They 
noted double-peaked profiles of the [\ion{N}{2}]$\lambda$6583 emission line 
inside the radius of $R\approx 6\arcsec$ which they treated as the simultaneous 
presence of the {\it corotating} and {\it counterrotating} gas at these radii.
However, the warm gas is a collisional dynamical subsystem, so two {\it coplanar} 
gaseous disks with opposite rotation senses cannot exist. 
Also, the [\ion{N}{2}]$\lambda$6583 line-of-sight velocity profile obtained 
by \citet{n3626hi} along the minor axis revealed deviations from zero level 
of $\pm 100$ km/s just inside $R\approx 6\arcsec$ (see their Fig.~7, bottom plot). 
It means that the secondary gas component detected by \citet{n3626hi} does not 
rotate circularly together with the stars in their symmetry plane. 
Furthermore, the careful inspection of the interferometric data
for the molecular gas CO(1-0) presented by \citet{n3626co} which have the spatial
resolution of $3.6\arcsec \times 2.9\arcsec$ has revealed that \citet{n3626co}
have also noticed the traces of this circumnuclear gas subsystem. They report
a presence of `an anomalous corotating gas component' which gives about 2.5\%\
of all molecular-gas emission. They called this component `corotating' because
they saw its contribution at the PV (position-velocity) diagram taken along
the major axis: in the radius range of $R=0\arcsec -4\arcsec$ it demonstrated the
motions `corotating` with the stellar ones, with the amplitudes of $\Delta v=180$~km/s --
see their Fig.~6, top. But over even larger extension, up to $R=8\arcsec - 10\arcsec$, 
this anomalous component is seen at the PV-diagram along the {\it minor} axis, and with
the same amplitude of $\Delta v=180$~km/s (their Fig.~6, bottom) -- 
just as the `orthogonally rotating' component of [\ion{O}{3}]$\lambda$5007 
in our Fig.~5, right-bottom.
\citet{n3626co} explained their finding of the `anomalous gas motions' by possible
strong non-circular motions -- namely, by inflow through the circumnuclear spiral 
density wave. But we think that the gas inflow with the velocity of more than 200 km/s
in the galaxy lacking a noticeable active nucleus seems to be incredible. Our
interpretation outlined below would be perhaps more appropriate.

\begin{figure*}
\plottwo{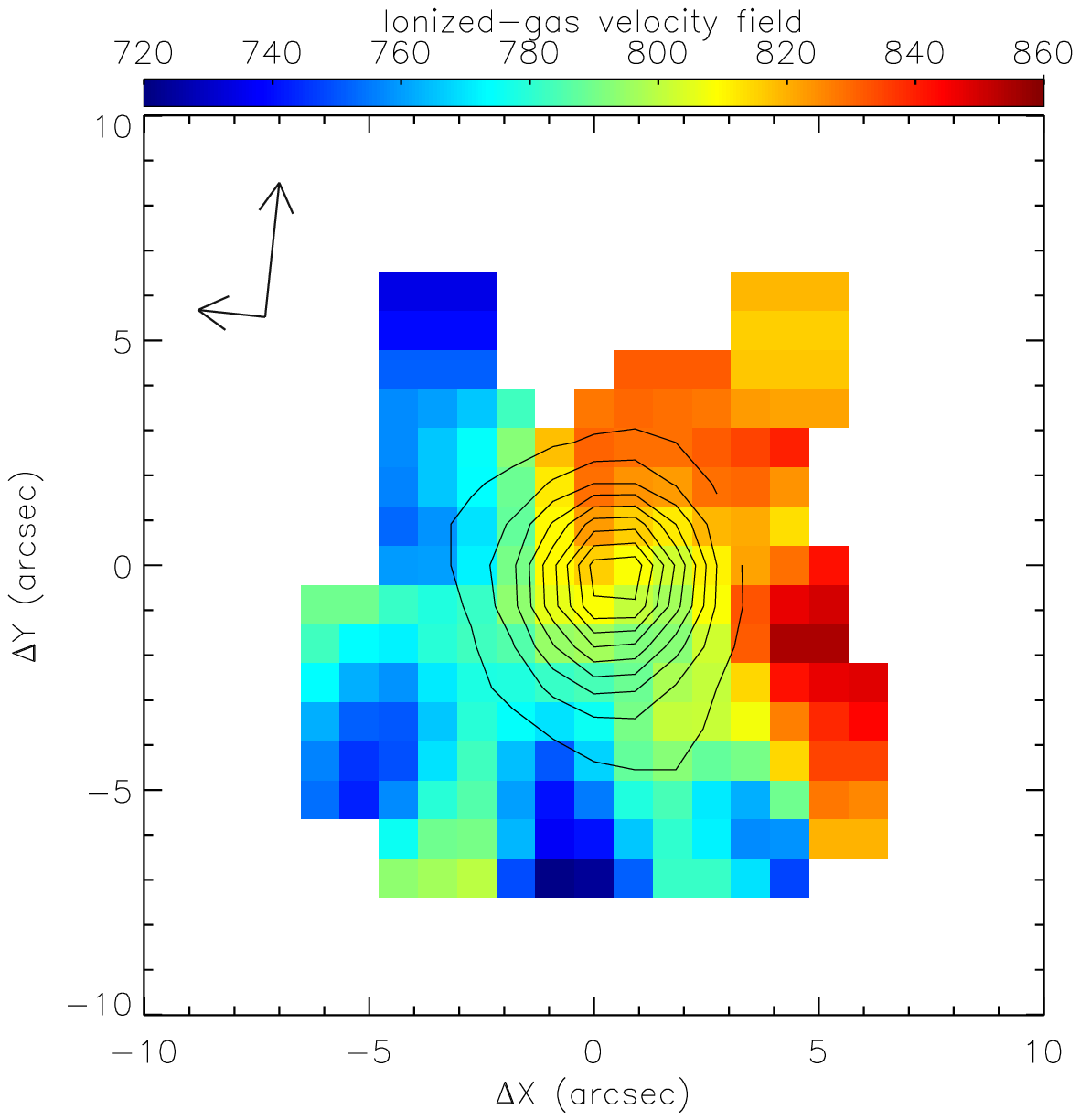}{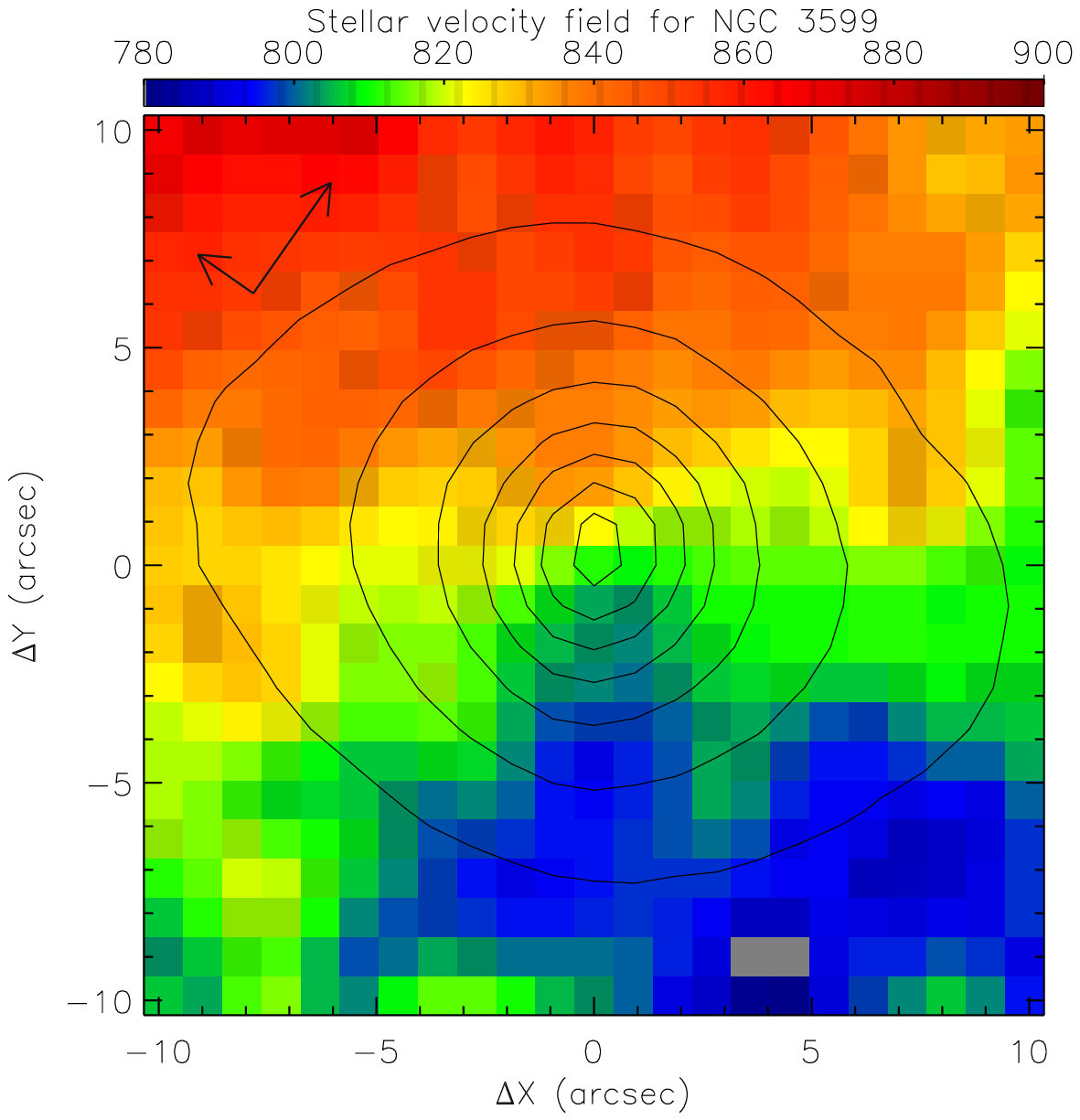}
\caption{The line-of-sight (LOS) velocity fields for
NGC~3599: {\bf left} -- the LOS velocity field of the ionized gas, from the
MPFS, {\bf right} -- the LOS velocity field of the stars, from the SAURON. 
The long arrow is directed to the north, the short one -- to the east.}
\end{figure*}

In Fig.~6 we combine together the data on photometric and kinematical major axes,
both for the stars and for the ionized gas of NGC~3626. The stellar velocity field
reveals a presence of separate circumnuclear stellar component; the tilted-ring
analysis shows that its kinematical orientation is close to the orientation
of the outer large-scale stellar disk:
$PA_0(\mbox{kin}) =341\degr \pm 5\degr$ and $i=32\degr \pm 6\degr$.
The slight discrepance of the kinematical inclinations between the outer disk
and the circumnuclear stellar region, keeping in mind their similar isophote
ellipticities, may imply that the latter is in fact a rather hot bulge.
At $R\approx 8\arcsec -12\arcsec$ the stellar kinematical major axis deviates
from the outer line of nodes; this deviation is due only to a couple of peculiar
spots on the stellar velocity field which are related to the high velocity
dispersion regions and so perhaps to `accreted' stars superposed onto the main
stellar component. Otherwise, the kinematical major axis of stars is quite
stable at the outer line of nodes that is very strange because the isophotes
within the same radius range have the other orientation. The situation looks 
the same as in NGC~3599. But the gas behavior in NGC~3626 is even more strange.
The gas kinematical major axis turns together with the continuum isophote
major axis following all its waving
pattern; in the very center they both reach $PA\approx 190\degr$. If we use
the $PA_0$ and $i$ values provided by the DETKA for the gas rotation plane
orientation and confront them to the orientation parameters of the stellar disk,
we obtain two solutions for the angle $\Delta i$ between two planes: it is $58\degr$ or
$87\degr$. In other words, at the radius $R < 4\arcsec-5\arcsec$ (400--500 pc)
we observe perhaps an inner gas polar ring. At larger radii, $R>7\arcsec$,
the bulk gas kinematical major axis stabilizes around $PA=170\degr -175\degr$ which
agrees perfectly with the orientation and rotation of the outer neutral-hydrogen
ring beyond the optical body of the galaxy \citep{n3626hi}; it is not surprising
because the outer neutral and the intermediate-radius warm gas have the same
rotation sense, counterrotating with respect to the stars, and the both gaseous
subsystems are obviously related by their origin.

\begin{figure*}
\plotone{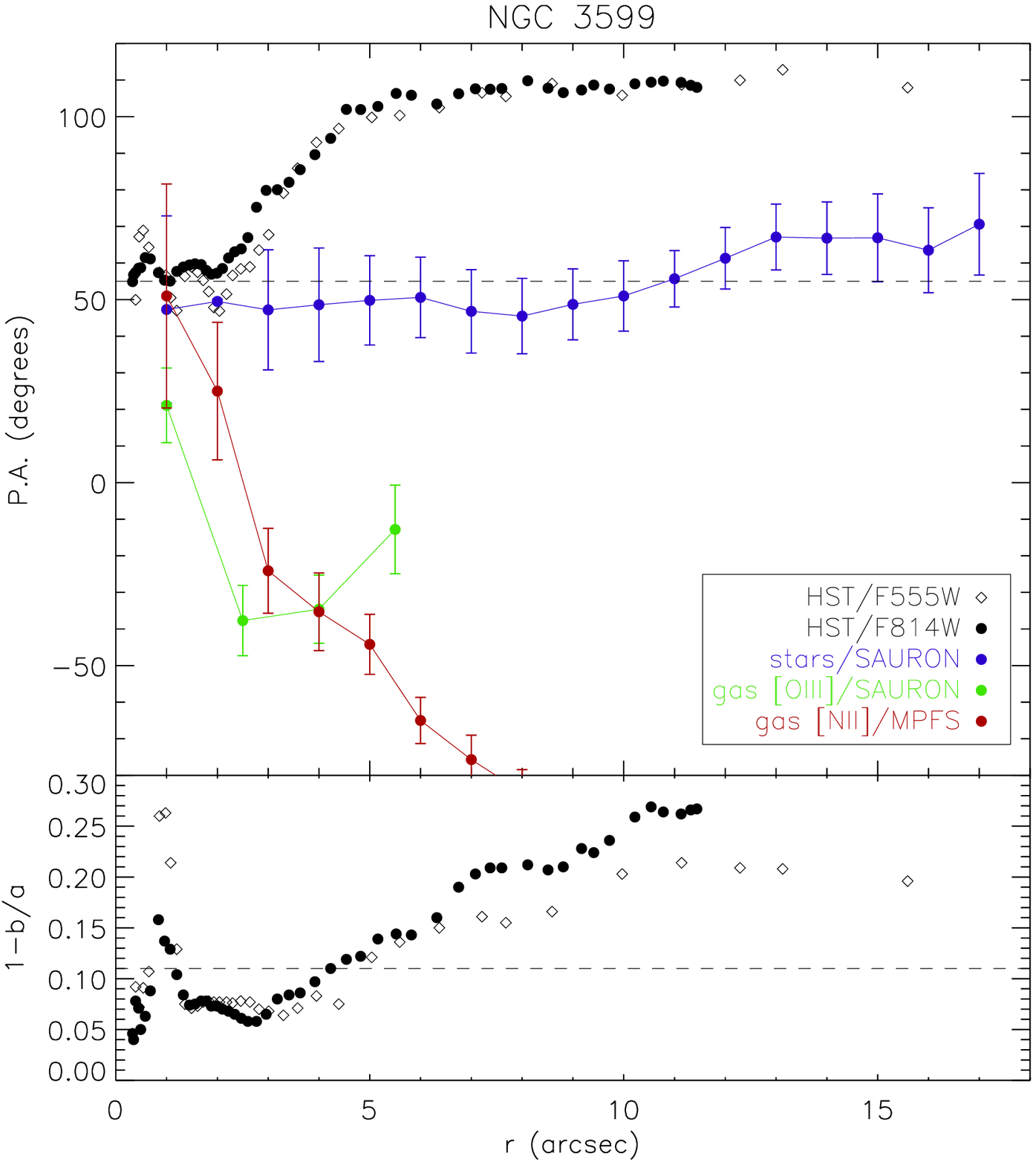}
\caption{The comparison of the photometric and kinematical major axes
in the central part of NGC~3599 ({\it top}), together with the high-resolution
isophote ellipticity profile ({\it bottom}).}
\end{figure*}

\begin{figure*}
\plottwo{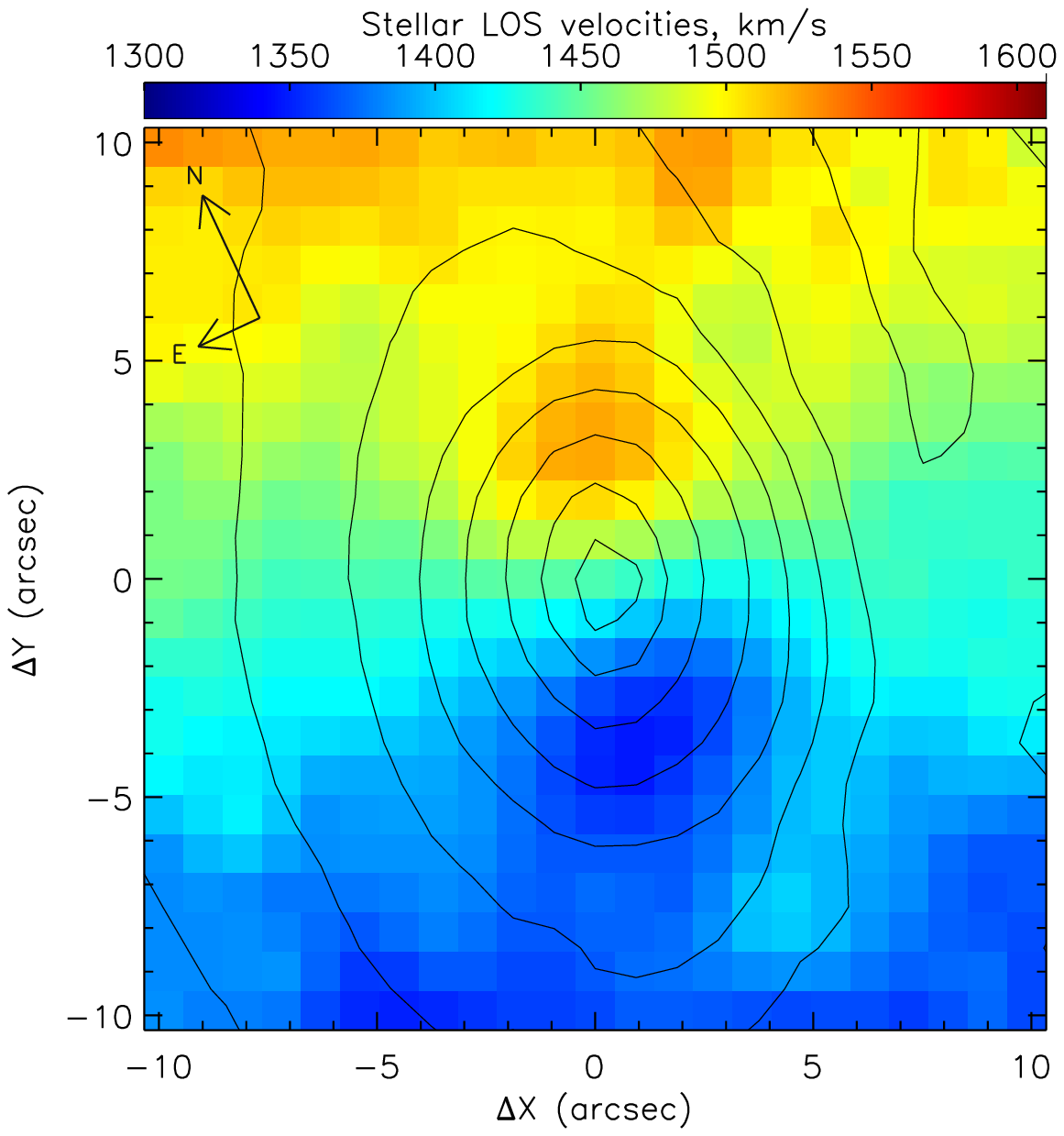}{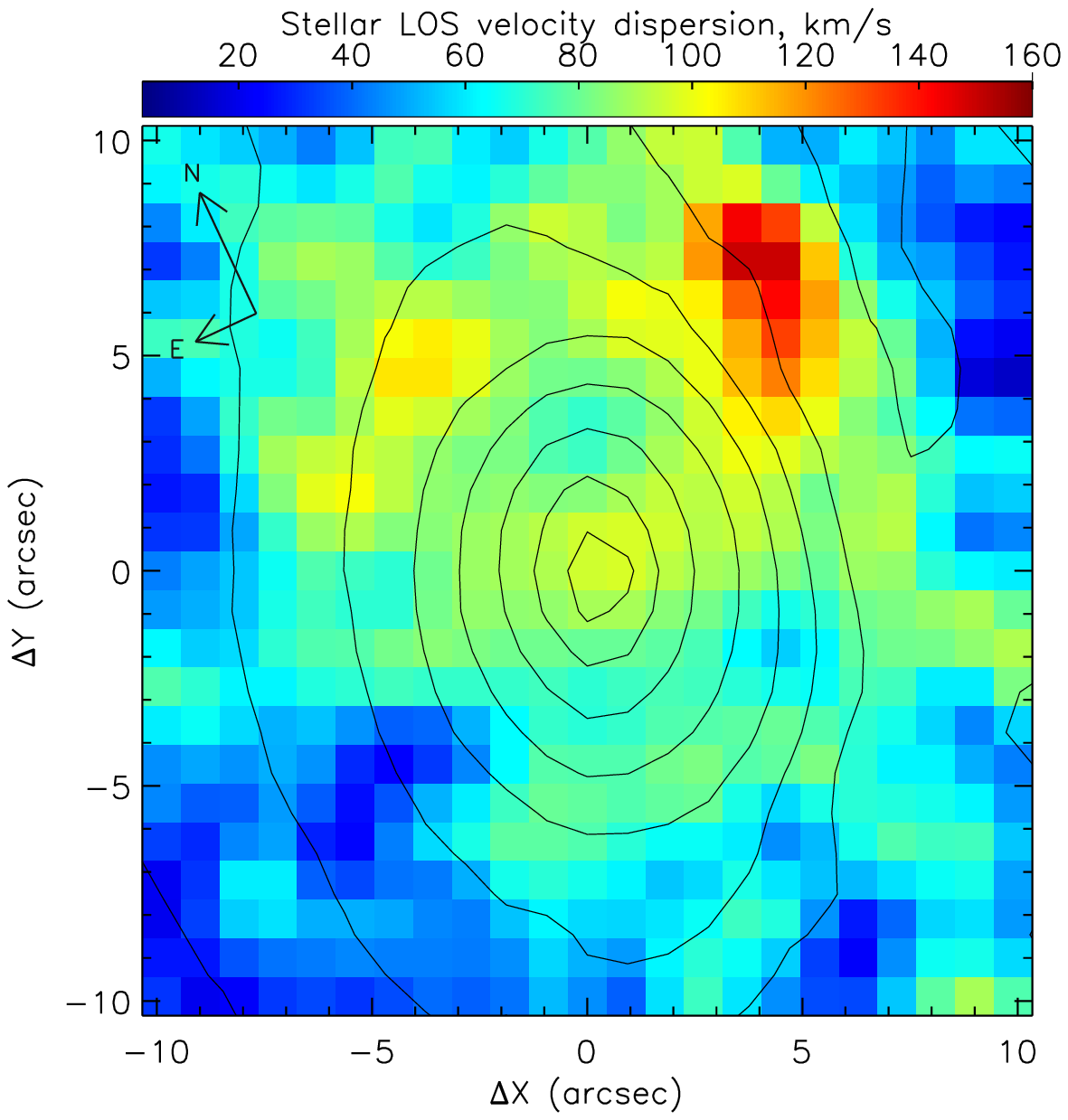}
\plottwo{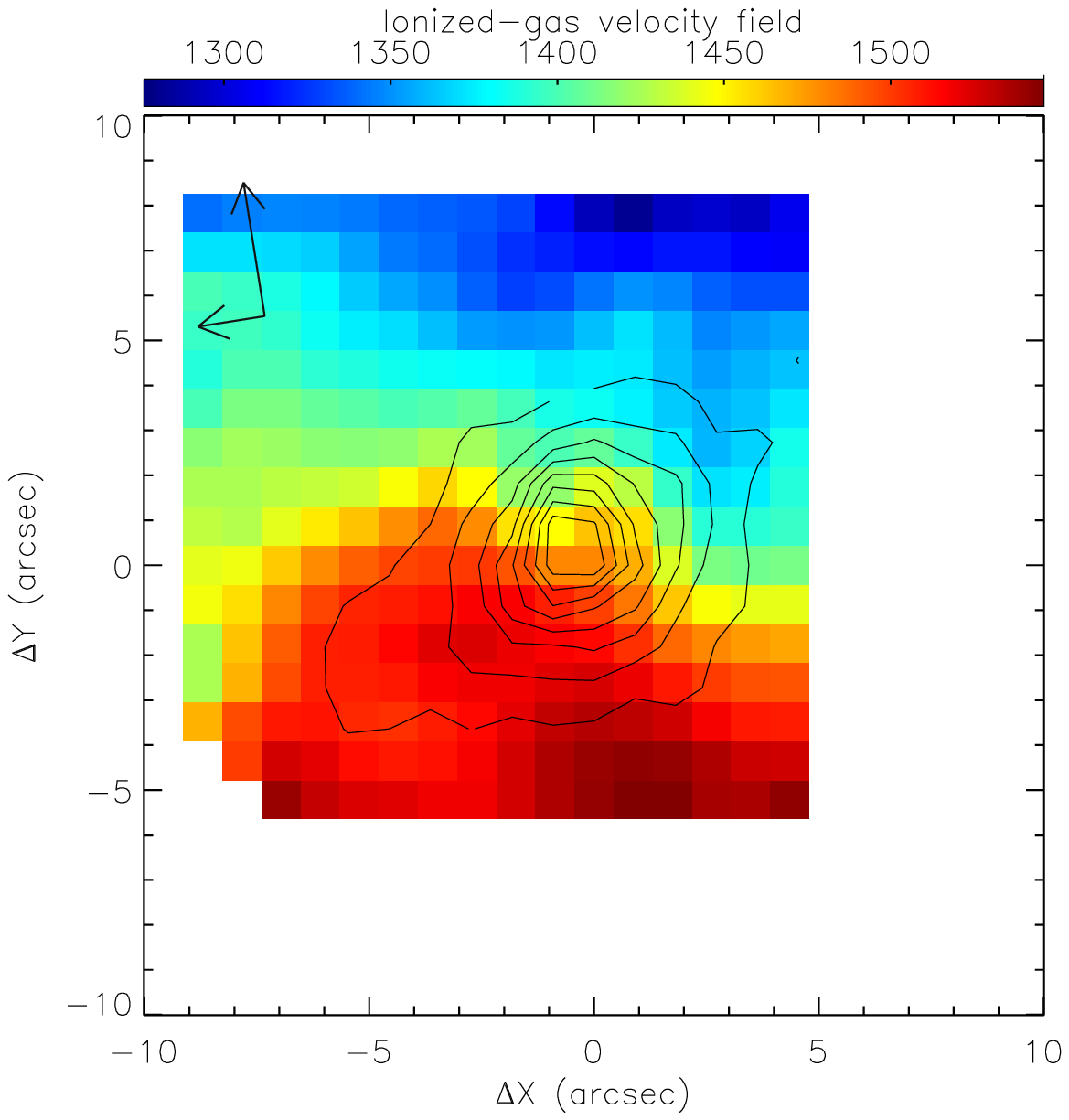}{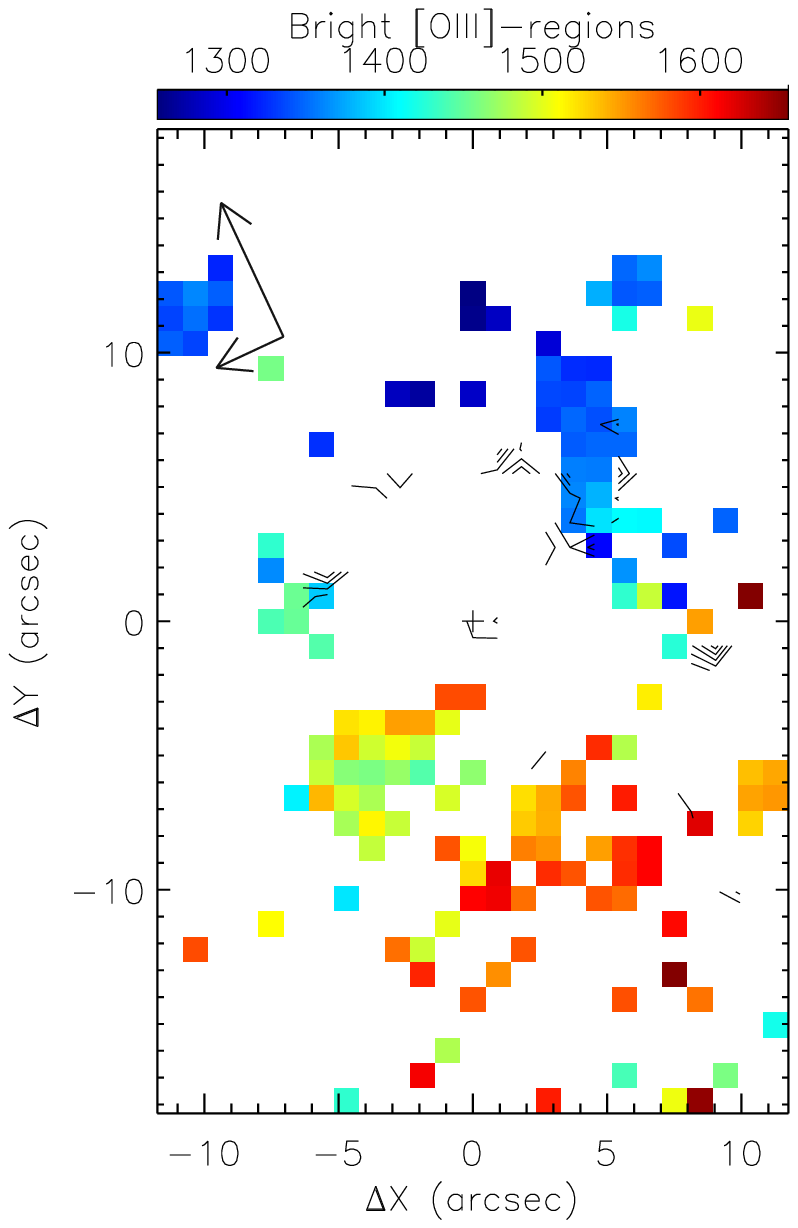}
\caption[f5anew.ps,f5bnew.ps,f5c.ps,f5d.ps]{The line-of-sight (LOS) velocity
fields for NGC~3626: {\bf top left} -- the LOS velocity field of the stars, 
from the SAURON, {\bf top right} -- the stellar LOS velocity dispersion, from
the SAURON, {\bf bottom left} -- the LOS velocity field of the ionized gas 
by the [\ion{N}{2}] emission line, from the MPFS, {\bf bottom right} -- the same, 
by the [\ion{O}{3}] emission line from the SAURON. The long arrow is directed 
to the north, the short one -- to the east.}
\end{figure*}

\section{Stellar populations and star formation in the centers of
NGC 3599 and NGC 3626.}

The Lick index system is well-calibrated on the MPFS data \citep{lenssum}
due to the wide spectral range and regular observations of the standard
Lick stars under the standard spectrograph configuration. The SAURON spectral
range is narrower than that of the MPFS; so there exist some problems
with the absorption-line index calibrations \citep{sauindex}.
We accept that the SAURON data are good to estimate qualitative index
distributions over the galaxy area under consideration, while the MPFS data
are good to made quantitative estimates of the stellar popupation properties
in the very centers of the galaxies.

Figure~7 presents the index-index diagrams for NGC~3599 and NGC~3626 where we compare
our MPFS data with the Single Stellar Population (SSP) models by \citet{thomod}.
Figures~7, top, demonstrate the fair solar magnesium-to-iron ratio in both 
galaxies, and this ratio does not vary along the radius within $8\arcsec $ from the 
centers. In the framework of the modern chemical evolution models, it means that the
duration of the main, or of the last, star formation epoch in the central regions
of the galaxies was longer than at least 1 Gyr everywhere including the nuclei.
Figure~7, bottom, confronting the H$\beta $-index to the combined metal-line index allows 
to estimate simultaneously the luminosity-weighted metallicity and age of the stellar
populations. Here two galaxies look somewhat different though the ages of their
nuclei coincides: they are 1-1.5~Gyr old or slightly younger. The environments of the
young stellar nucleus in NGC~3599 are evidently older:
at $R > 3\arcsec $ the SSP age rises to 5 Gyr at least, and
the metallicity drops from several times solar to the subsolar one.
In NGC~3626 the SSP age stays at the level of 1 Gyr up to
the radial distance of $\sim 7\arcsec $ where the emission-line intensities 
increase dramatically, and the age estimates become somewhat uncertain.

In Fig.~7, bottom, we have plotted the H$\beta$ indices corrected for the modest
emission contamination.
We have calculated the corrections by using our red-range spectroscopic data of
the MPFS. Our approach is based on the fact that the H$\alpha $ emission line
is always much stronger than the H$\beta $ emission line while the equivalent
widths of Balmer {\bf absorption} lines are comparable, and for the intermediate-age
stellar populations the H$\alpha $ absorption is even weaker than the higher-order
lines \citep{balsil}. Moreover, the Balmer emission-line intensity
ratios are narrowly fixed by the mechanism of gas excitation; for example, the lowest
ratio $I(\mbox{H}\alpha )/I(\mbox{H}\beta )$, 2.5, is for the gas excitation by young
massive stars \citep{burgess}, while shock excitation and excitation by power-law
continuum (by AGN) give higher $I(\mbox{H}\alpha ) /I(\mbox{H}\beta )$ ratio. So we
have measured the equivalent widths of the H$\alpha$ emission lines in the MPFS
red spectra co-added over rings centered onto the galactic nuclei and have obtained
$EW(\mbox{H}\alpha emis)$ as a function of the distance from the centers. Then, to
obtain the H$\beta$ index corrections for the nuclei of NGC~3599 and NGC~3626,
we have divided the measured $EW(\mbox{H}\alpha emis)$ by 4 following the empirical
prescription by \citet{sts2001} for the combined mechanisms of gas
excitation. For the region of NGC~3626 at $R\ge 4\arcsec $ we have used the
coefficient of 2.7 because it is a site of the ongoing star formation as we
shall see below. Over the off-nuclear region of NGC~3599 the emission lines
are weak, and the corrections are applied only within $R\le 2\arcsec$.

\begin{figure*}
\plotone{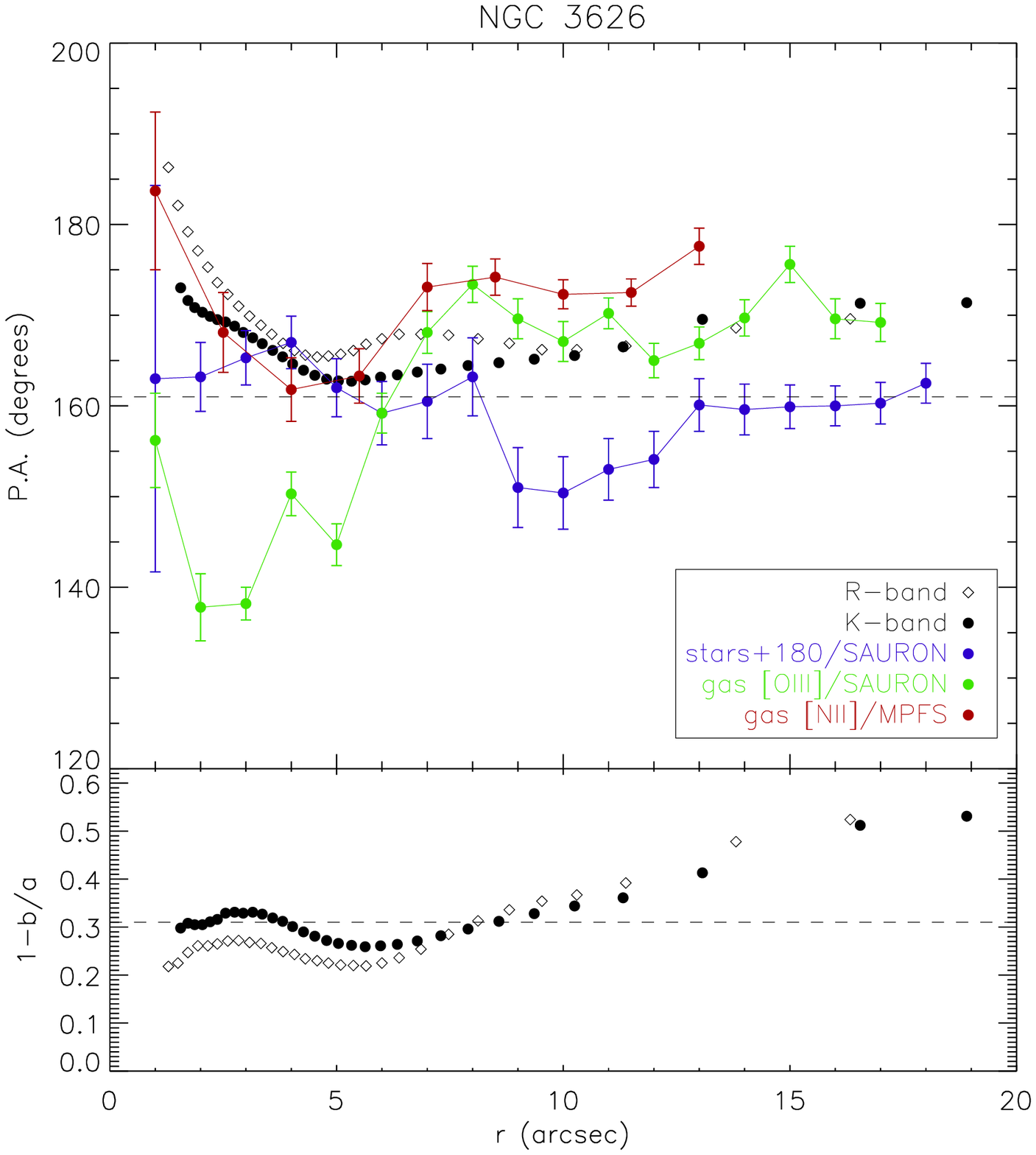}
\caption{The comparison of the photometric and kinematical major
axes in the central part of NGC~3626 ({\it top}), together with the
isophote ellipticity profile ({\it bottom}).}
\end{figure*}

\begin{figure*}
\plottwo{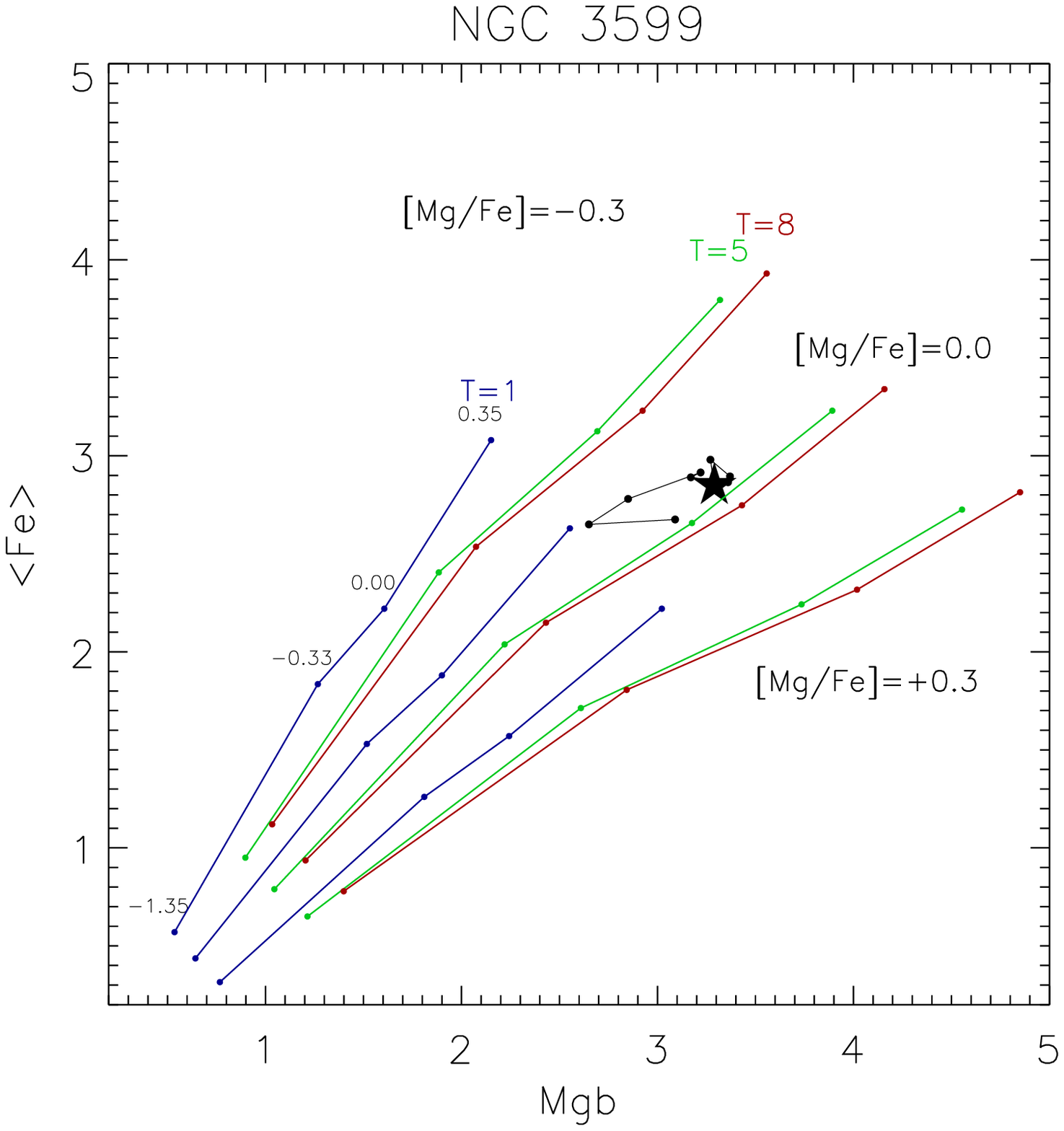}{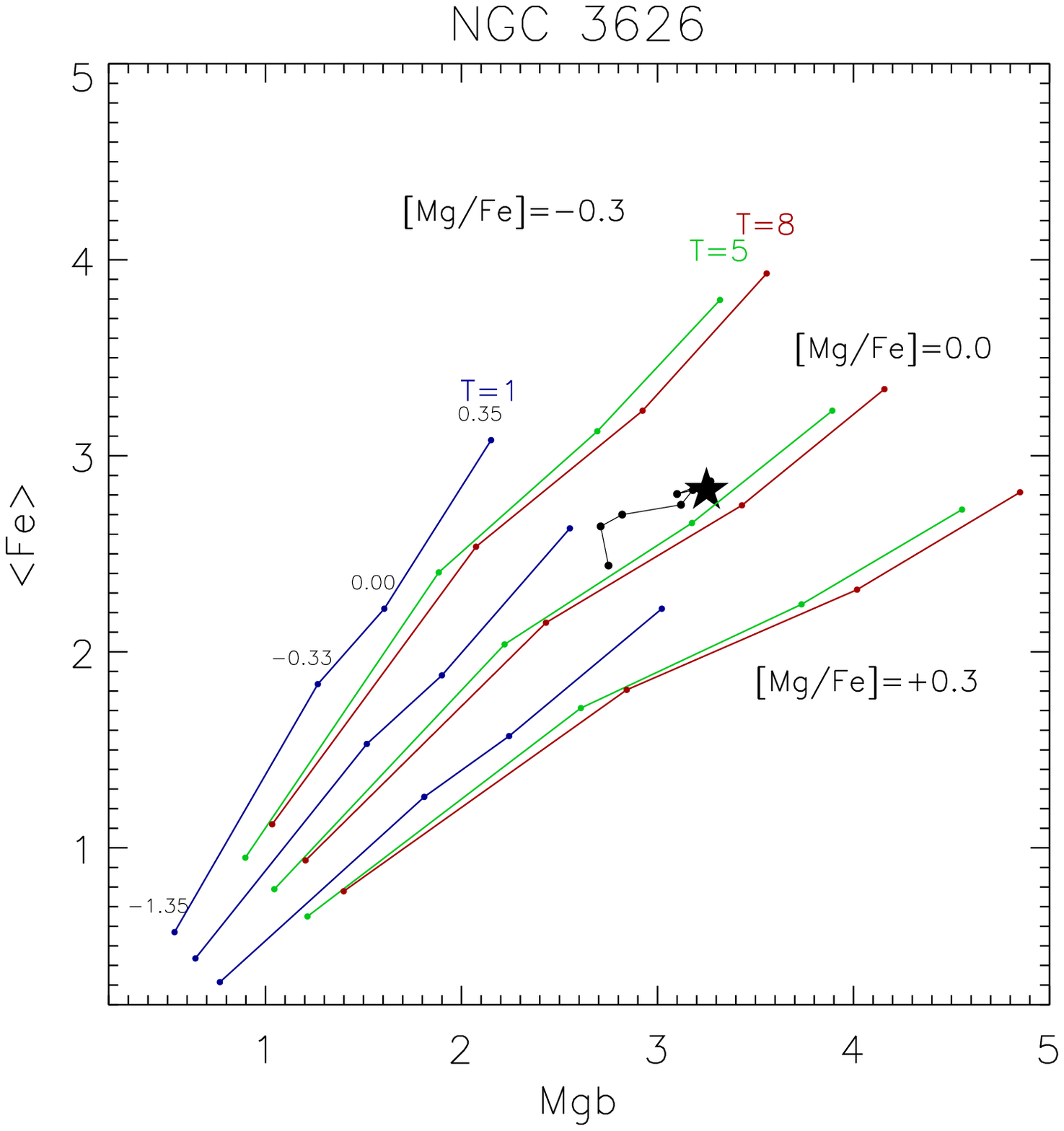}
\epsscale{0.5}
\plotone{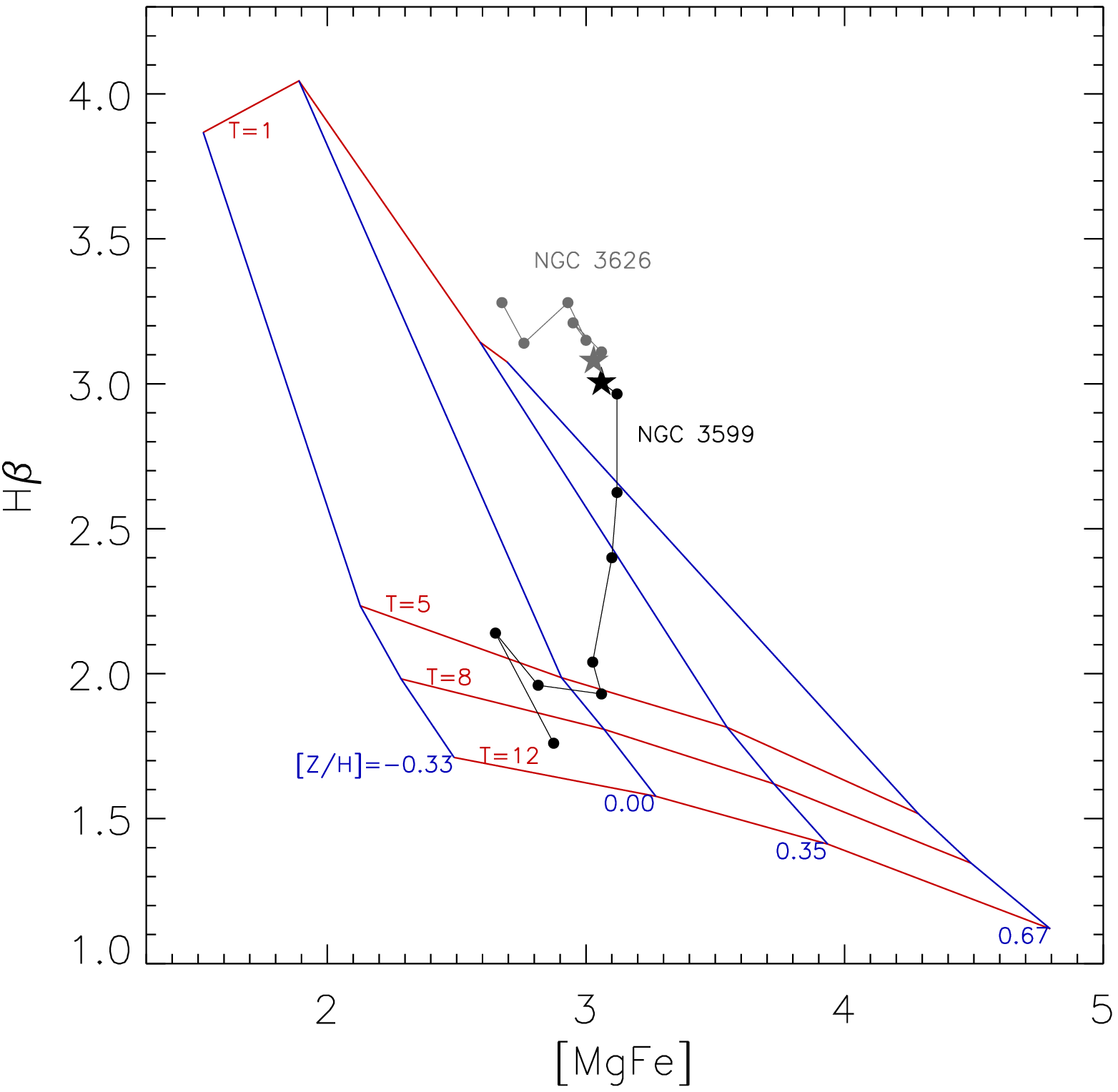}
\caption{{\bf top} --
The $\langle \mbox{Fe} \rangle$ vs Mgb diagrams for
NGC~3599 and NGC~3626 azimuthally averaged index measurements. The black
points are taken with the radial step of 1 arcsec, starting from the nuclei 
plotted by the large stars.
The accuracy of the ring-averaged indices is 0.1~\AA--0.15~\AA. The simple
stellar population models by Thomas et al. (2003) for three different
magnesium-to-iron ratios (--0.3, 0.0, and $+0.3$) and three different ages 
(1, 5, and 8 Gyr) are plotted as reference. The small signs
along the model curves mark the metallicities of +0.35, 0.00,
--0.33, and --1.35, if one takes the signs from
right to left. {\bf bottom} -- The age-diagnostic diagram for the stellar
populations in the central parts of the galaxies under consideration;
the H$\beta$-index measurements are rectified
from the emission contamination as described in the text.
The typical accuracy of the indices is 0.1~\AA\ for the combined metal-line index
and 0.15~\AA\ for the H$\beta$. The stellar population models by
\citet{thomod} for [Mg/Fe]$=0.0$
and four different ages (1, 5, 8, and 12 Gyr, from top to bottom curves)
are plotted as reference frame; the dashed lines crossing the
model curves mark the metallicities of +0.67, +0.35, 0.00,
--0.33 from right to left.}
\end{figure*}

Forcing by the necessity to correct the H$\beta $ index, we have analyzed the
gas excitation mechanisms for the central regions of NGC~3599 and NGC~3626
in the manner of  \citet{vo}, by calculating the emission-line
intensity ratios. By measuring the SCORPIO spectra for NGC~3599, we have applied
the Gauss analysis to the combination of the H$\alpha $ absorption and emission
lines; to measure [\ion{O}{3}]$\lambda $5007 at the bottom of the \ion{Ti}{1}$\lambda $5007
absorption line we have used the SAURON spectra co-added in rings. The results are
astonishing. The nucleus of the dwarf lenticular galaxy NGC~3599 demonstrates
$I(\mbox{[OIII]})/I(\mbox{H}\beta ) > 3$,
$I(\mbox{[NII]}\lambda 6584)/ I(\mbox{H}\alpha ) = 2.15$, and
$I(\mbox{[SII]}\lambda 6717+6730)/ I(\mbox{H}\alpha ) = 0.48$; in other words,
according to the modern criteria by \citet{kewley06} (see their Fig.4, middle),
the nucleus of NGC~3599 is classified as a Seyfert nucleus. It is the second evidence
for the presence of a supermassive black hole in the center of this dwarf galaxy,
after the discovery of the X-ray burst by \citet{n3599xray}. Within the radius of
$2\arcsec $ from the center of NGC~3599 where the emission lines are still
well-measurable, their ratios remain to be far from the criteria for the excitation
by young stars and give evidences for the AGN excitation or shock excitation.
We do not see any signs of star formation in the emission-line spectra of NGC~3599.
But the color map of the galaxy provided by the HST imaging (Fig.~8) reveals
a spectacular blue ring, with the radius of $1\farcs 7$ and the color difference
of about $\Delta (V-I) \approx 0.25$ with respect to the galaxy center.
Since we do not see HII-type excitation, the age of the starburst in the ring
must be more than 10 Myr.

NGC~3626 has also rings of young stars in the central part. The $B-V$ color
map (Fig.~9 left) demonstrates inhomogeneous oval blue zone, with the bluest points,
$B-V \approx 0.7 -0.75$, in $3\arcsec $ to the north-east and in $5\arcsec $ to
the north-west. The mean stellar age within this zone, according to the Fig.~7,
is about 1 Gyr. However just beyond the outer radius of this blue ring the
intense H$\alpha $ emission arises, and at $R=7\arcsec $ to the north we see
an emission-line ratio $I(\mbox{[NII]}\lambda 6584)/ I(\mbox{H}\alpha )
\approx  0.4$ (Fig.~9 right) that is a net evidence for the current star formation
at this radius \citep{kewley06}. Interestingly, the color of {\it this}
ring is not blue, it is even red to the west from the nucleus (Fig.~9 left).
Evidently, the starforming ring at the outer edge of the bulge of NGC~3626
is buried by a large amount of dust. It is a little surprising that the blue
knot is still seen quite clearly in $3\arcsec $ to the east of the nucleus, and
the more distant HII-region (a part of the starforming ring) is hidden by the
bulge and is not seen. Perhaps, warping of the gaseous disk plane in the center
of the galaxy that has been revealed by the kinematical data (Fig.~6) is a
real geometrical distortion.

To see in more detail the history of star formation in the center of NGC~3626,
we present the Lick index maps derived from the SAURON data in Fig.~10 and from
the MPFS data in Fig.~11. One
can see immediately that the maps contain a lot of compact features which
have been eliminated by azimuthal averaging in Fig.~7. At the H$\beta $ map
{\bf not} correcting for the emission, a ring of intense current star formation
is outstanding by the negative values of the measured index. However, besides this
outstanding feature, we note a prominent peak of the H$\beta $ absorption-line
index in $4\arcsec - 5\arcsec $ to the north-west of the nucleus, close to
one of the bluest spots in the $(B-V)$-color map of Fig.~9 (left). We have found a
counterpart to this feature at the Mgb map -- it is a local minimum of the
magnesium index. The analysis of the stellar population properties of this
local `spot' with the MPFS data at the index-index diagrams reveals that it 
does not differ by an age but is somewhat less metal-rich than its surroundings. 
Taking in mind a considerable counterrotating gas content of NGC~3626, we can
speculate that near the nucleus we may see a remnant of the merged dwarf
galaxy, more exactly, the densest part of it which has spiraled through
the whole disk of NGC~3626 during a few Gyrs after merging. Then the spot
of the high stellar velocity dispersion somewhat farther from the center
may be the precessing, less dense stellar `tail' of the merged satellite.
The concentric elliptical rings of the young blue stars and of the H$\alpha $
emission tracing the current star formation may then be consequences of
the interaction between accreted counterrotating gas and the prograde
pre-existing gas of  NGC~3626, like in NGC~3593 \citep{corsini}. 
An alternative explanation of the ring structure similar to that
observed in NGC~3626 is proposed by numerical simulations of close passages 
of galaxies by \citet{Tutukov2006}. The concentric stellar  rings are 
formed when a companion moves in the equatorial plane in the direction opposite 
to the rotation of the studied galaxy (see their Fig.~3). The  large-scale 
gas counterrotation observed in  NGC~3626 suggests just such geometry of the 
galactic interaction. Interestingly, the radial concentration of star formation 
in NGC~3626 shifts with time: the starforming ring expands slowly outward.

\begin{figure*}
\plotone{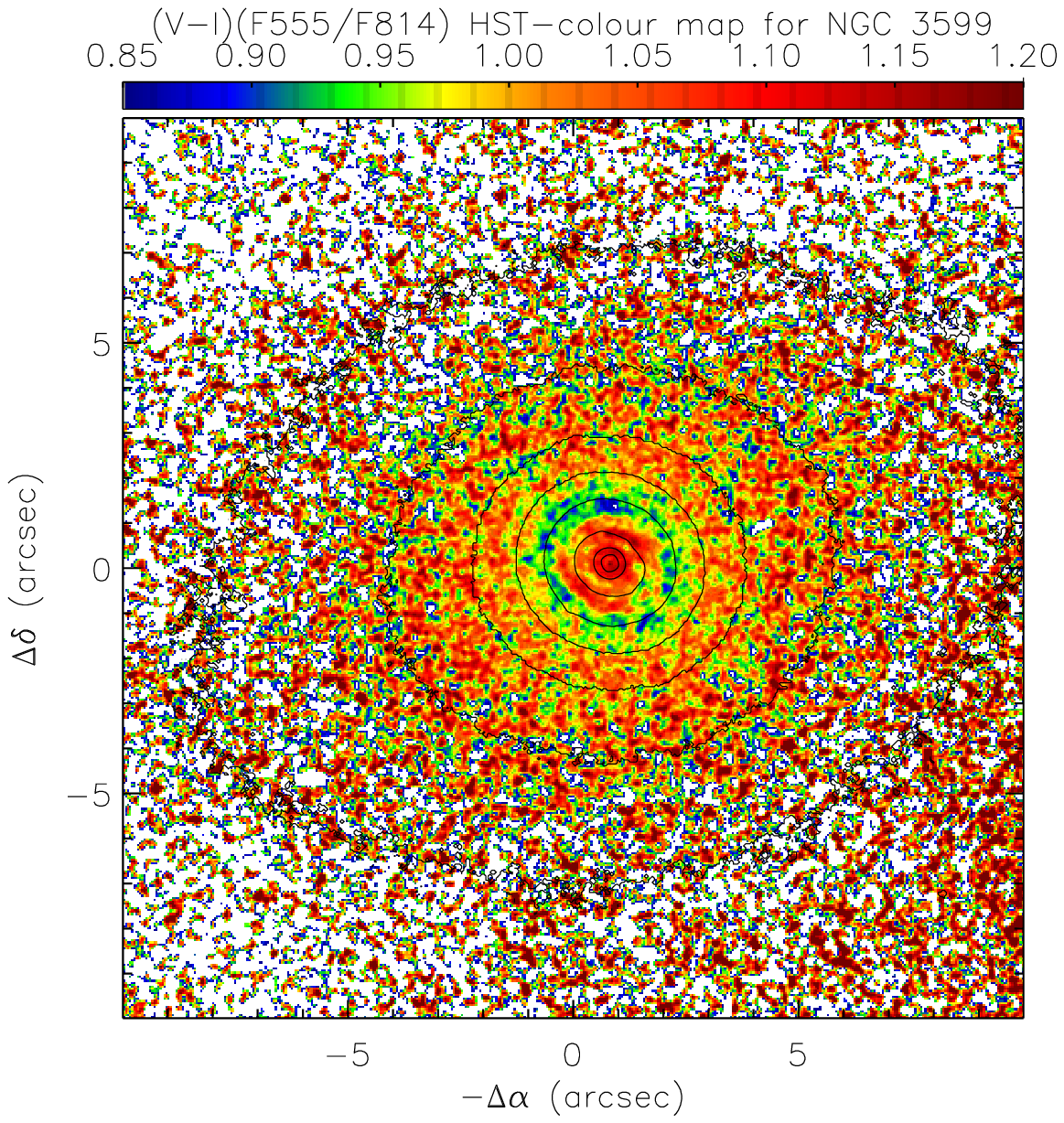}
\caption{The HST color map for the central part of NGC~3599, the isophotes
superposed are from the F814W-image.}
\end{figure*}

\begin{figure*}
\plottwo{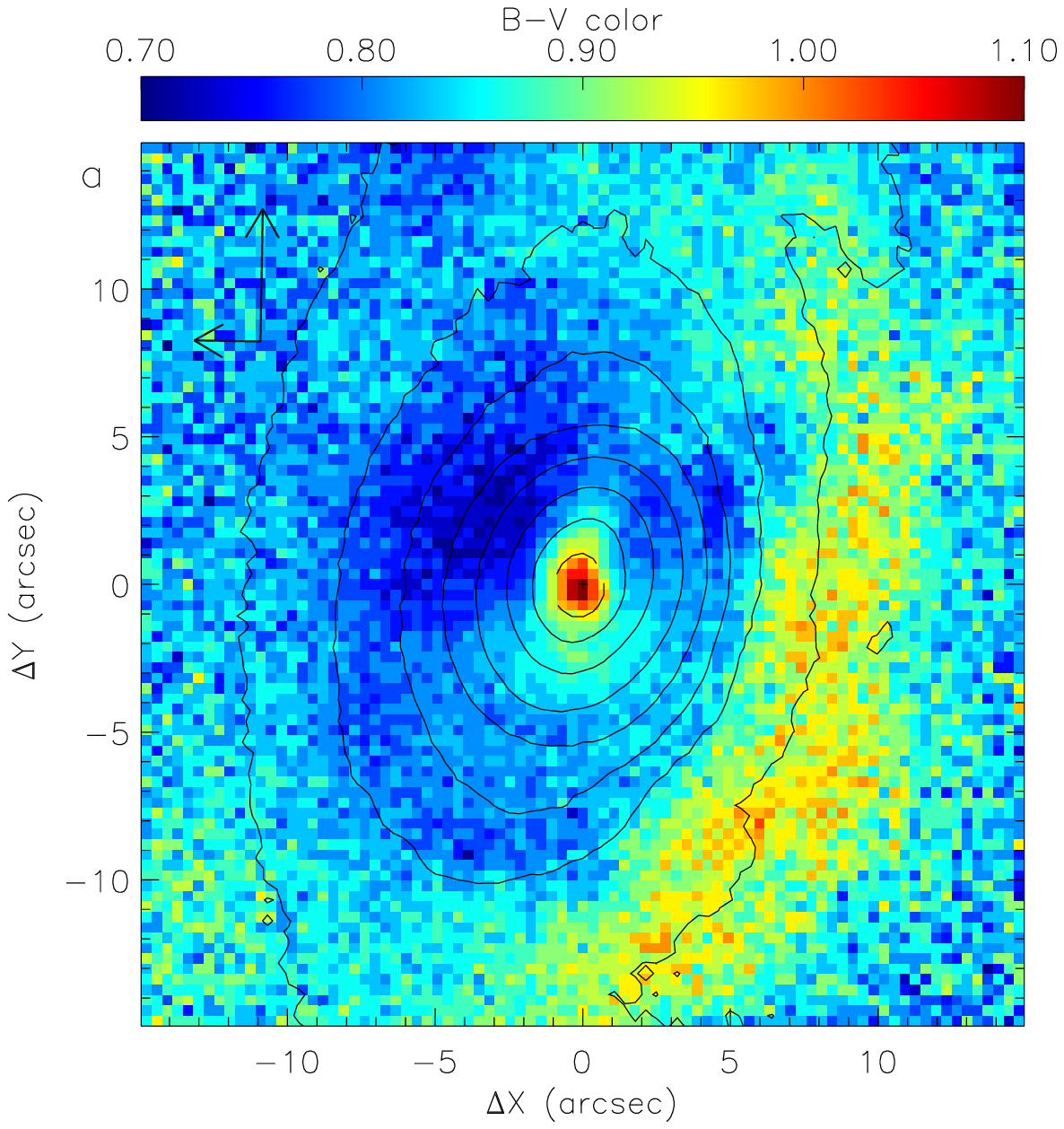}{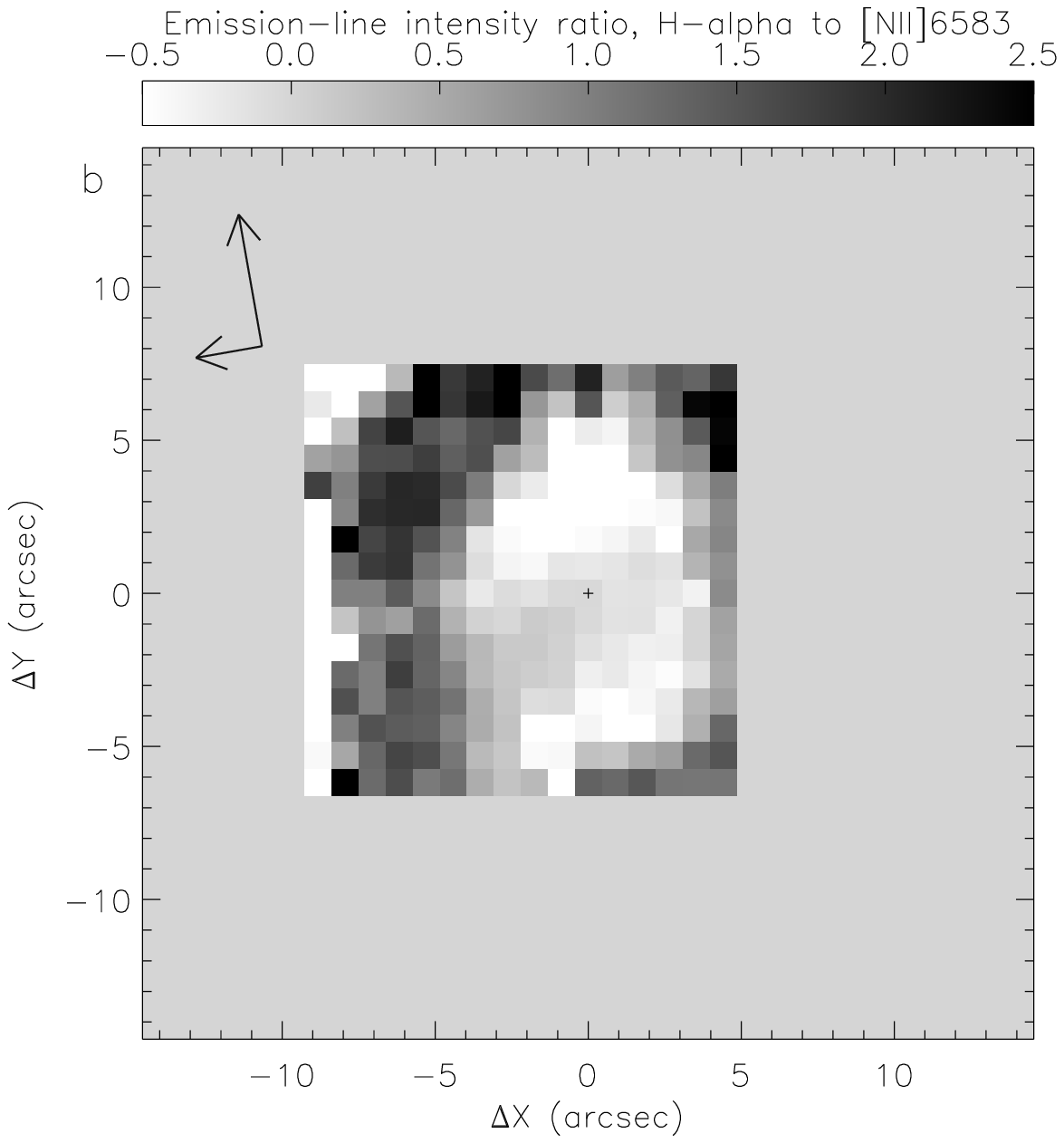}
\caption{The $B-V$ color map ({\bf left}) and the map
of the H$\alpha$ to [\ion{N}{2}]$\lambda 6584$ emission-line ratio ({\bf right})
for the central part of NGC~3626. The negative ratios mean that the H$\alpha$
is here an absorption line. The long arrow is directed to the north, the short one 
-- to the east.}
\end{figure*}

\section{Conclusions and brief discussion}

Both NGC~3599 and NGC~3626 are lenticular galaxies at the outskirts of
the X-ray bright group. They are located beyond the dense intragroup
medium area. In contrast to the `old' lenticular galaxy NGC~3607 -- the
central galaxy of the group, -- they are `young' lenticular galaxies: the
mean ages of the central stellar populations are less than 2 Gyr in both
galaxies. If we relate the age of the last star formation in the nuclei
to the galaxy transformation from spirals, these events occured rather
recently comparing to the dynamical timescale, and some other signs might be
still present and may help to identify these events. Indeed, both galaxies
demonstrate a complex of features evidencing for the recent minor merging.
First of all, the gas kinematics is quite decoupled from the stellar kinematics,
and we may suggest that the gas which is observed in the central parts
of the galaxies mostly has recently been acquired. Both galaxies look isolated
inside the group, so it is not mere accretion, it is just minor merging.
The gas rotation planes are strongly warped, and at some distances from the
center become almost polar with respect to the galactic large-scale disks. 
Namely, NGC~3626 hosts the inner polar  disk at the radii of $R<500$ pc, whereas 
in NGC~3599 the ionized gas comes to significantly titled orbits out of $R\sim 200$ pc.

Near the very center in NGC~3599 and beyond some radius
in NGC~3626, the gas rotation planes come closer to the main galactic planes,
and the presence of star-forming nuclear rings in both galaxies can perhaps
be explained by the accreted gas interaction with the own primordial gas of
the galaxies, as it is the case in NGC~3593 \citep{corsini} or Arp~212
\citep{moiseevarp}. Moreover, the geometry of the gaseous disk in NGC~3599 looks 
just like a picture observed in Arp~212: the orbits of gas clouds are coplanar 
with a stellar disk in the circumnuclear regions and become warped (may be polar) 
at larger distances from the center. First of all, the compressed primordial gas 
has to be exhausted by the recent star formation, and the accreted counterrotating
gas is left to be observed. Later, the counterrotating gas would be also
reprocessed into stars, and we will see a counterrotating stellar component
as is now seen in NGC~3593 \citep{n3593bertola}.

NGC~3626, a large galaxy, has a huge amount of extended HI gas,
while NGC~3599, the dwarf, has not. The extended gaseous disk of NGC~3626
counterrotates the stars, and if their rotation planes are coplanar, we
can remember the simulations of an isolated stellar-gaseous disk by
\citet{sec1} where initially counterrotating gas begins to inflow
because of the bar and forms a stable highly inclined ring near the center
due to vertical resonance effects. Though both galaxies are classified as
SA (non-barred), their large-scale structures reveal multi-tiered stellar
disks, and the inner disks in both galaxies are evidently oval. So we may
suggest that the bars were present some time ago, but are now dissolved, or
almost dissolved; dynamical simulations demonstrate that dissolving bars
leave weakly oval stellar disks after their dissolution \citep{bshj06}. The 
current starforming ring in NGC~3626 may be then related to the inner Lindblad 
resonance of this dissolving bar.

The whole scenario of the galaxy transformations from  spirals into lenticulars
may include the following sequence of related events: firstly, minor merger,
then the acquired gas and stars inflow to the center, the large-scale stellar
disks develop bars and are heated, the gas is compressed near the center and
exhausted by the intense induced star formation. After a few Gyrs, we have
typical lenticular galaxies, only with the unrelaxed gas behavior. 
Within such scenario, the main mechanism of the lenticular galaxy
formation is gravitational and not related at all to the intragroup hot
medium impact.

\begin{figure*}
\plotone{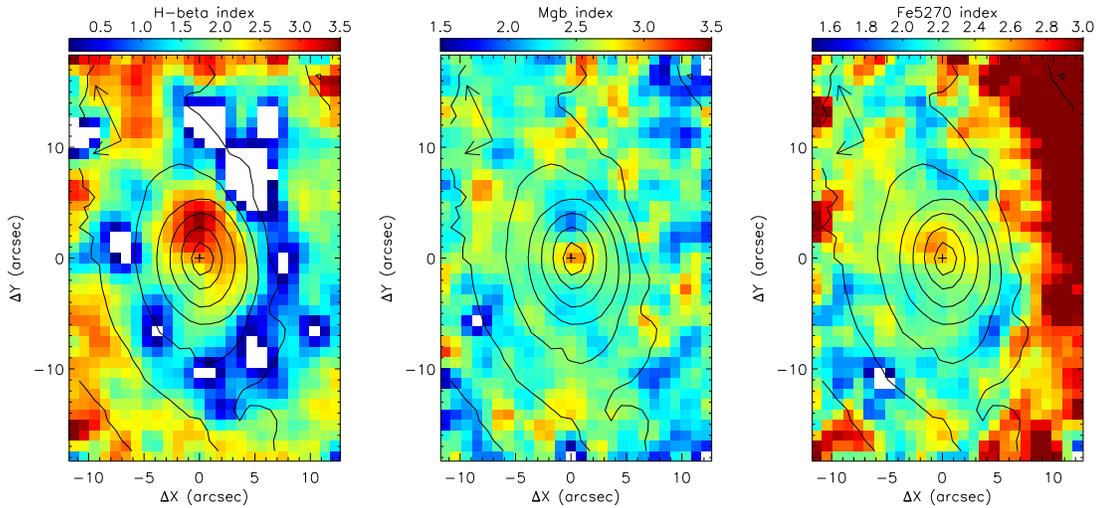}
\caption{The Lick index maps for the central part of
NGC~3626 derived from the SAURON data: 
{\bf left} -- H$\beta$, {\bf middle} -- Mgb,  {\bf right} -- Fe5270. The long
arrow is directed to the north, the short one -- to the east.}
\end{figure*}

\begin{figure*}
\plotone{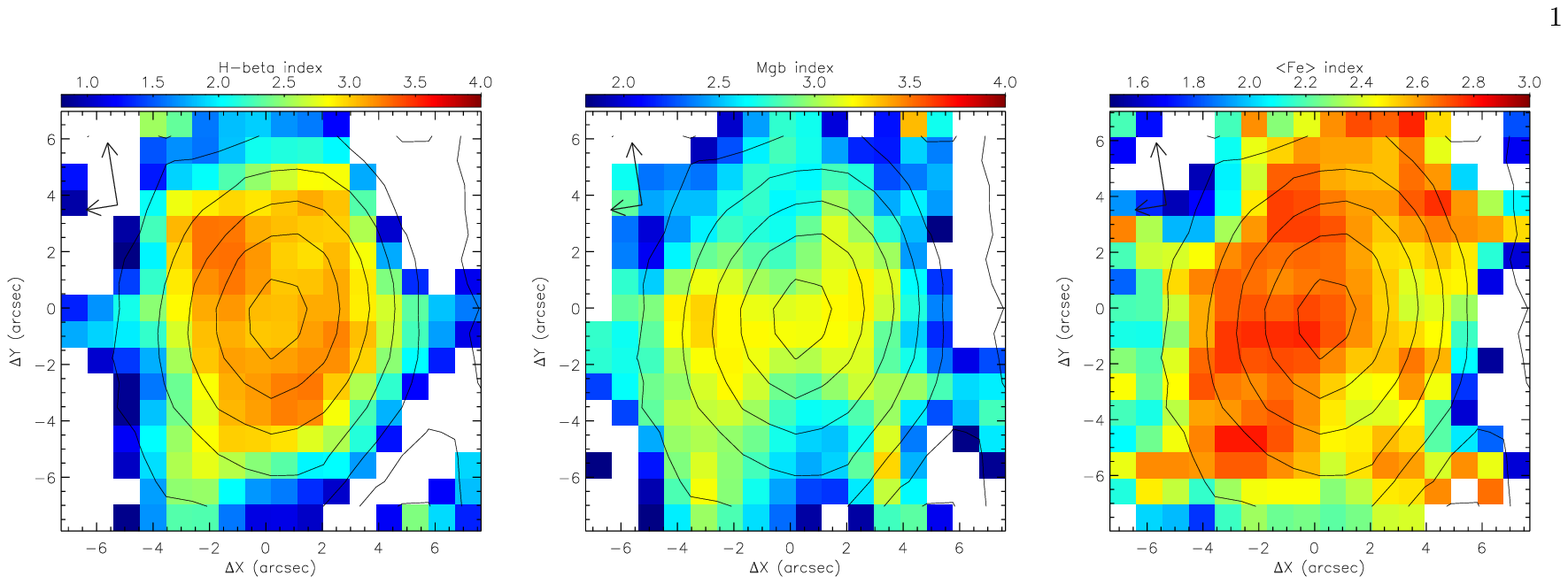}
\caption{The Lick index maps for the central part of
NGC~3626 derived from the MPFS data: 
{\bf  left} -- H$\beta$,   {\bf middle}  -- Mgb,  {\bf  right} -- $\langle \mbox{Fe} \rangle$. 
The long arrow is directed to the north, the short one -- to the east.}
\end{figure*}

\acknowledgements

We thank Prof. V. L. Afanasiev for supporting the Multi-Pupil Fiber
Spectrograph of the 6m telescope and for taking part in some of the
observations which data are used in this work. The 6m telescope is
operated under the financial support of the Science Ministry of Russia
(registration number 01-43). During our data analysis
we used the Lyon-Meudon Extragalactic Database (LEDA) supplied by the
LEDA team at the CRAL-Observatoire de Lyon (France) and the NASA/IPAC
Extragalactic Database (NED) operated by the Jet Propulsion
Laboratory, California Institute of Technology under contract with
the National Aeronautics and Space Administration.
This research is partly based on data obtained from the Isaak Newton
Group Archive which is maintained as part of the CASU Astronomical
Data Centre at the Institute of Astronomy, Cambridge,
on observations made with the NASA/ESA Hubble Space Telescope, obtained
from the data archive at the Space Telescope Science Institute, which is
operated by the Association of Universities for Research in Astronomy,
Inc., under NASA contract NAS 5-26555, and on SDSS data.
Funding for the Sloan Digital Sky Survey (SDSS) and SDSS-II has been
provided by the Alfred P. Sloan Foundation, the Participating Institutions,
the National Science Foundation, the U.S. Department of Energy, the National
Aeronautics and Space Administration, the Japanese Monbukagakusho,
and the Max Planck Society, and the Higher Education Funding Council
for England. The SDSS Web site is http://www.sdss.org/.
The work on the study of multi-tiered disk galaxies
is supported by the grants of the Russian Foundation for Basic
Researches number 07-02-00229a and number 09-02-00870a.


\begin{thebibliography}{}



\bibitem[Adelman-McCarthy et al.(2008)]{sdssdr6} Adelman-McCarthy, J.,
     Agueros, M.A., Allam, S.S., et al. 2008, \apjs, 175, 297



\bibitem[Afanasiev \& Moiseev(2005)]{scorpref} Afanasiev, V.L., Moiseev, A.V.,
2005, Astronomy Letters,  31, 193  (astro-ph/0502095)



\bibitem[Afanasiev \& Sil'chenko(2007)]{leo2cen} Afanasiev, V.L., Sil'chenko,
O.K., 2007, Astron. Astrophys. Trans., 26, 317



\bibitem[Afanasiev et al.(2001)]{mpfsref} Afanasiev, V.L., Dodonov, S.N.,
  Moiseev, A.V., 2001, In: Stellar dynamics: from classic to modern/
  Eds. Osipkov L.P. and Nikiforov I.I., Saint Petersburg Univ. press, 103



\bibitem[Bacon et al.(1995)]{betal95} Bacon, R., Adam, G., Baranne, A.,
    Courtes, G., Bubet, D., et al. 1995, \aaps, 113, 347



\bibitem[Bacon et al.(2001)]{betal01} Bacon, R., Copin, Y., Monnet, G.,
   Miller, B.W., Allington-Smith, J.R., Bureau, M., Carollo, C.M., Davies,
   R.L.,  Emsellem, E., Kuntschner, H., Peletier, R.F., Verolme, E.K.,
   de Zeeuw, P.T., 2001, \mnras, 326, 23



\bibitem[Balinskaya \& Sil'chenko(1993)]{balsil} Balinskaya, I.S., Sil'chenko,
O.K. 1993, Bull. of SAO, 35, 43



\bibitem[Berentzen et al.(2006)]{bshj06} Berentzen, I., Shlosman, I.,
Jogee, S. 2006, \apj, 637, 582



\bibitem[Bertola et al.(1996)]{n3593bertola} Bertola, F., Cinzano, P., Corsini,
E.M., Pizzella, A., Persic, M., Salucci, P. 1996, \apj, 458, L67



\bibitem[Burgess(1958)]{burgess} Burgess, A., 1958, \mnras, 118, 477



\bibitem[Ciri et al.(1995)]{n3626ciri} Ciri, R., Bettoni, D., Galletta, G. 1995,
Nature, 375, 661



\bibitem[Corsini et al.(1998)]{corsini} Corsini, E.M., Pizzella, A., Funes, J.G.,
Vega Beltran, J.C., Bertola F. 1998, \aap, 337, 80



\bibitem[de Souza et al.(2004)]{budda} de Souza, R.E., Gadotti, D.A.,
dos Anjos, S. 2004, \apjs, 153, 411



\bibitem[Dressler et al.(1997)]{dressler} Dressler, A., Oemler, A., Jr.,
Couch, W.J., Smail, I., Ellis, R.S., et al. 1997, \apj, 490, 577



\bibitem[Elmegreen \& Elmegreen(1985)]{eebars85} Elmegreen, B.G., Elmegreen, D.M.
 1985, \apj, 288, 438



\bibitem[Elmegreen et al.(1996)]{eebars96} Elmegreen, B.G., Elmegreen, D.M.,
Chromey, F.R., Hasselbacher, D.A., Bissell, B.A. 1996, \aj, 111, 2233



\bibitem[Erwin et al.(2008)]{erwinbars} Erwin, P., Pohlen, M., Beckman, J.E. 2008,
     \aj, 135, 20



\bibitem[Esquej et al.(2008)]{n3599xray} Esquej, P., Saxton, R. D., Komossa, S.,
    Read, A. M., Freyberg, M. J. et al. 2008, \aap, 489, 543



\bibitem[Fasano et al.(2000)]{fasano} Fasano, G., Poggianti, B.M., Couch, W.J.,
     Bettoni, D., Kjaergaard, P., Moles, M., 2000, \apj,  542,  673



\bibitem[Friedli \& Benz(1993)]{sec1} Friedli, D., Benz, W. 1993,
         \aap, 268, 65



\bibitem[Garcia-Burillo et al.(1998)]{n3626co} Garcia-Burillo, S., Sempere, M. J.,
Bettoni, D. 1998, \apj, 502, 235



\bibitem[Haynes et al.(2000)]{n3626hi} Haynes, M.P., Jore, K. P., Barrett, E. A.,
  Broeils, A. H., Murray, B. M. 2000, \aj, 120, 703



\bibitem[Just et al.(2010)]{just2010} Just, D.W., Zaritsky, D., Sand, D.J., Desai, V.,
Rudnick, G. 2010, \apj, 711, 192



\bibitem[Kewley et al.(2006)]{kewley06} Kewley, L.J., Groves, B., Kauffmann, G.,
   Heckman T. 2006, \mnras, 372, 961



\bibitem[Kuntschner et al.(2006)]{sauindex} Kuntschner, H., Emsellem, E., Bacon,
   R., Bureau, M., Cappellari, M., et al. 2006, \mnras, 369, 497



\bibitem[Laurikainen et al.(2005)]{lauri05} Laurikainen, E., Salo, H., Buta, R. 2005,
\mnras, 362, 1319



\bibitem[Mahdavi et al.(2000)]{xraygroup} Mahdavi, A., Bohringer, H.,
   Geller, M., Ramella, M. 2000, \apj, 534, 114



\bibitem[Magrelli et al.(1992)]{magrelli} Magrelli, G., Bettoni D., Galletta G. 1992,
\mnras, 256, 500



\bibitem[Moiseev(2008)]{moiseevarp} Moiseev, A.V. 2008, Astrophys. Bull., 63, 201



\bibitem[Moiseev et al.(2004)]{moiseevdb} Moiseev, A.V., Valdes, J.-R.,
      Chavushyan V.H. 2004, \aap, 421, 433



\bibitem[Mulchaey et al.(2002)]{xrayleo2} Mulchaey, J.S., Davis, D.S., Mushotzky,
R.F., Burstein, D. 2002, \apjs, 145, 39



\bibitem[Pohlen \& Trujillo(2006)]{sdsslate} Pohlen, M., Trujillo, I. 2006,
   \aap, 454, 759



\bibitem[Ramella et al.(1997)]{catgroup} Ramella, A., Pisani, A.,
  Geller, M.J. 1997, \aj, 113, 483



\bibitem[Sil'chenko(1997)]{sil97} Sil'chenko O.K., 1997, Astronomy Reports, 41, 567



\bibitem[Sil'chenko(2006)]{lenssum} Sil'chenko, O.K., 2006, \apj, 641, 229



\bibitem[Sil'chenko \& Afanasiev(2008)]{n80mpfs} Sil'chenko, O.K., Afanasiev,
     V.L., 2008, Astronomy Reports, 52, 799



\bibitem[Sil'chenko et al.(2002)]{s0_3} Sil'chenko, O.K., Afanasiev, V.L.,
    Chavushyan V.H., Valdes J.-R., 2002, \apj, 577, 668



\bibitem[Sil'chenko et al.(2003)]{we2003} Sil'chenko, O.K., Moiseev, A.V.,
    Afanasiev V.L., Chavushyan V.H., Valdes J.-R., 2003, \apj, 591, 185



\bibitem[Startseva et al.(2009)]{n80phot} Startseva, M.A.,
Sil'chenko, O.K., Moiseev, A.V. 2009, Astronomy Reports, 53, 1101



\bibitem[Stasinska \& Sodr\'e(2001)]{sts2001} Stasinska, G., Sodr\'e, I.,
     Jr. 2001, \aap, 374, 919



\bibitem[Thomas et al.(2003)]{thomod} Thomas, D., Maraston, C.,
      Bender, R. 2003, \mnras, 339, 897



\bibitem[Tonry et al.(2001)]{sbfdist} Tonry, J.L., Dressler, A.,
  Blakeslee, J.P., Ajhar, E.A., Fletcher, A.B., et al. 2001, \apj,
  546, 681

  

\bibitem[{Tutukov \& Fedorova(2006)}]{Tutukov2006}
{Tutukov}, A.~V., {Fedorova}, A.~V., 2006, Astronomy Reports 50, 785 



\bibitem[Veilleux \& Osterbrock(1987)]{vo} Veilleux, S., Osterbrock, D.E. 1987,
\apjs, 63, 295



\bibitem[Wilman et al.(2009)]{wilman} Wilman, D.J., Oemler, A., Mulchaey, J. S.,
McGee, S. L., Balogh, M. L., Bower, R. G. 2009, \apj, 692, 298




\bibitem[Worthey et al.(1994)]{woretal} Worthey, G., Faber, S.M.,
   Gonz\`alez, J.J., Burstein, D. 1994, \apjs, 94, 687



\bibitem[Wozniak et al.(1995)]{3bphot} Wozniak, H., Friedli, D., Martinet,
      L., Martin, P., Bratschi, P., 1995, \aaps, 111, 115



\end{thebibliography}
\end{document}